\documentclass[aps,cha,twocolumn,groupedaddress]{revtex4}
\usepackage{graphicx}
\usepackage{amsmath}

\begin{document}
\title{Network Structure Effects in Reservoir Computers}
\author{T. L. Carroll}
\email{thomas.carroll@nrl.navy.mil}
\affiliation{US Naval Research Lab, Washington, DC 20375}
\author{L. M. Pecora}
\email{louis.pecora@nrl.navy.mil}
\affiliation{US Naval Research Lab, Washington, DC 20375}

\date{\today}

\begin{abstract}
A reservoir computer is a complex nonlinear dynamical system that has been shown to be useful for solving certain problems, such as prediction of chaotic signals, speech recognition or control of robotic systems. Typically a reservoir computer is constructed by connecting a large number of nonlinear nodes in a network, driving the nodes with an input signal and using the node outputs to fit a training signal. In this work, we set up reservoirs where the edges (or connections) between all the network nodes are either +1 or 0, and proceed to alter the network structure by flipping some of these edges from +1 to -1. We use this simple network because it turns out to be easy to characterize; we may use the fraction of edges flipped as a measure of how much we have altered the network. In some cases, the network can be rearranged in a finite number of ways without changing its structure; these rearrangements are symmetries of the network, and the number of symmetries is also useful for characterizing the network. We find that changing the number of edges flipped in the network changes the rank of the covariance of a matrix consisting of the time series from the different nodes in the network, and speculate that this rank is important for understanding the reservoir computer performance.
\end{abstract}

\maketitle

{\bf A reservoir computer is a high dimensional dynamical system that may be used for computations on time series signals. Usually the dynamical system is created by connecting a number of nonlinear nodes in a network that includes feedback. A time series input signal is coupled into the nodes of the reservoir computer and the time series output of the individual nodes are recorded. To train the reservoir computer, the node output time series are fit to some training time series that is related to the input signal, often with a least squares fit. The network itself is not changed. The fitting coefficients contain information about the relationship between the input signal and the training signal. We study several ways to characterize the structure of the network to see how the network structure affects the error in fitting the training signal. We find that if we choose a simple network where connections (or edges) between the nodes are all +1 or zero, we can produce simple changes in the network by flipping some of the edges to -1. For some networks, the network with edges flipped may contain symmetries; that is, the nodes may be rearranged without altering the network structure. In that case, the network may be characterized by the number of these symmetries. Otherwise, the network can be characterized by the fraction of edges flipped from +1 to -1. We hope that by understanding how the network structure affects the performance of the reservoir computer, we can build reservoir computers that produce more accurate solutions to problems. }

\section{Introduction}
Reservoir computers were developed as a type of recurrent neural network by machine learning researchers \cite{jaeger2001,natschlaeger2002} , but they may also be described using the language of dynamical systems. An advantage of reservoir computers over other machine learning techniques is that training a reservoir computer is fast and simple; the reservoir itself, a network of nonlinear nodes, is kept fixed, and the time series responses from the nodes are combined using a linear weighted sum, where the weights are varied to fit a training signal. 

Reservoir computers have been shown to be useful for solving a number of problems, including reconstruction and prediction of chaotic attractors \cite{lu2018,zimmerman2018,antonik2018,lu2017,jaeger2004}, recognizing speech, handwriting or other images \cite{jalavand2018} or controlling robotic systems \cite{lukosevicius2012}. One attractive feature of reservoir computers is that they may be implemented in a wide range of analog hardware, making them potentially very fast but with low power consumption. Examples of reservoir computers so far include photonic systems \cite{larger2012, van_der_sande2017}, analog circuits \cite{schurmann2004}, mechanical systems \cite{dion2018} and  field programmable gate arrays \cite{canaday2018}.

One obstacle to understanding what reservoir computers can or can't do is that there is only a limited amount of theory on how reservoir computers function.  Much of the theoretical work hinges on understanding the tradeoff between nonlinearity in the reservoir computer nodes and memory \cite{jaeger2002,inubushi2017, marzen2017}, where memory is described by the time dependent correlation between the reservoir computer input and the reservoir computer fit to delayed versions of this input signal. Other work focuses on generalized synchronization \cite{lu2018, lymburn2019}. The previous work doesn't describe how the choice of the network influences the reservoir computer, so in this work we study the effect of different networks on the reservoir computer.

Early on, Jaeger \cite{jaeger2002} stated "If the network is suitably inhomogeneous, the various echo functions will significantly differ from each other". It does seem obvious that a diverse network is necessary for an efficient reservoir computer, but how can this inhomogeneity be measured? A standard practice is to use a sparse random network. Random networks are easy to achieve in simulations, but the options for creating a network in a real analog system may be limited by experimental constraints. It would be interesting to confirm the standard wisdom using some measure of inhomogeneity which could then be used to aid in the design of a network in a system where a creating completely random network might not be possible.

Characterizing a random network is difficult, so we use a network that is simple to characterize. All of the connections between nodes in our networks have the values $\pm 1$ or 0. Within these limits, the connections are chosen randomly. All networks are connected, which means for any two nodes, there is a path between them.

The random networks are initialized with all edges set to +1 or 0, and then some of the edges are flipped from +1  to -1.We will see that if we take two copies of the same initial network and flip the same number of edges but choose different edges to flip, the performance of the reservoir computer may be different. Therefore we choose many networks where the same number of elements are flipped and track trends in the behavior of the entire group of reservoir computers.

We will examine two different node types and two different input signals. The nodes are based on a nonlinear differential equation or a leaky hyperbolic tangent map \cite{jaeger2002}, while the input signals will come from a Lorenz chaotic system or a nonlinear map acting on a random signal. The nonlinear mapping was chosen from a set of problems commonly used to test reservoir computers \cite{rodan2011}.

In section \ref{computers}, we will describe reservoir computers and show how they may be trained. Section \ref{input} describes our choice of how to feed signals into the reservoir computer, while section \ref{adjacency} discusses how the network adjacency matrix may be characterized. All elements of the adjacency matrix, which describes the edges between nodes,  start as +1 or 0, and some of the +1 edges are flipped to -1. The values of the edges are then normalized so that the maximum of the absolute value of the real part of the set of eigenvalues of the adjacency matrix is 0.5. We may then characterize the adjacency matrix by the number of symmetries it contains or by the fraction of the elements flipped from +1 to -1. 

The main method we use to characterize reservoir computer performance is the testing error, described in section \ref{computers}, but we also use other methods to analyze the reservoir networks. These methods are described in section \ref{analysis}. Section \ref{symmetry} describes how symmetries in the network may be found, section \ref{cov_rank} lays out the computation of the rank of the covariance matrix for the reservoir, and section \ref{mem_cal} summarizes a common method for calculating memory capacity.

After laying out the analysis methods, section \ref{signals} describes how we create the input signals from a Lorenz chaotic system or a random signal system. We then proceed to simulations of the reservoir computer testing error as we flip network edges in section \ref{simulations}. After first showing how the choice of input vector affects our results in section \ref{net_comp},  we study the reservoir computer training error as a function of the number of symmetries in the network (section \ref{sym_sim}) and as the fraction of edges flipped (section \ref{flip}). This section also includes measurements of the covariance matrix rank. The memory capacity for different networks is calculated in section \ref{mem_cap}. Finally, both fraction flipped and sparsity are varied in section \ref{sparse_var}.

\section{Reservoir Computers}
\label{computers}
\begin{figure}[ht]
\centering
\includegraphics[scale=0.4]{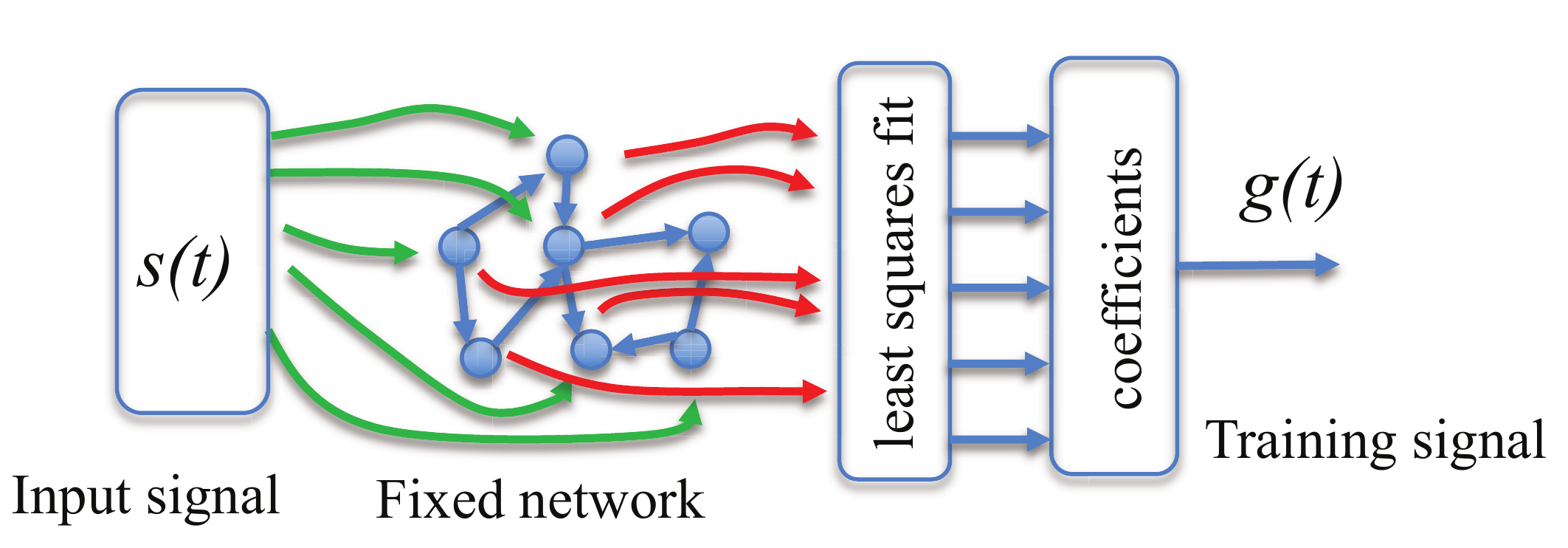} 
  \caption{ \label{reservoir_computer}
Block diagram of a reservoir computer. We have an input signal $s(t)$ that we want to analyze, and a related training signal $g(t)$. When trained, the reservoir computer will estimate $g(t)$ from $s(t)$. In the training phase, the input signal $s(t)$ drives a fixed network of nonlinear nodes, and the time varying signals from the nodes are fit to the training signal $g(t)$ by a least squares fit. The coefficients are the result of the training phase. To use the reservoir computer for computation, a different signal $s'(t)$ is input to the reservoir computer. As an example, \cite{lu2017}, $s(t)$ was a Lorenz $x$ signal, while $s'(t)$ was a Lorenz $x$ signal started with different initial conditions. The time varying node signals that result from $s'(t)$ are multiplied by the coefficients from the training phase to produce the output signal $g'(t)$, which in \cite{lu2017} was a good fit to the Lorenz $z$ signal corresponding to $s'(t)$.}
  \end{figure}

We used a reservoir computer to estimate one time series signal based on a different (but related) time series signal.
Figure \ref{reservoir_computer} is a block diagram of a reservoir computer. There is an input signal $s(t)$ from which the goal is to extract information, and a training signal $g(t)$ which is used to train the reservoir computer. In \cite{lu2017} for example, $s(t)$ was the $x$ signal from a Lorenz chaotic system, while $g(t)$ was the Lorenz $z$ signal. The reservoir computer was trained to estimate the $z$ signal from the $x$ signal.

There are no specific requirements on the nodes in a reservoir computer, other than when all nodes are connected into a network, the network should be stable; that is, it should settle into a stable fixed point. Commonly used nodes include hyperbolic tangent \cite{manjunath2013} or sigmoid functions \cite{verstraeten2010}, but in analog experiments the node nonlinearity is determined by the experimental system \cite{larger2012, van_der_sande2017,schurmann2004,dion2018,canaday2018}. We select two different node types to decrease the chance that our results depend on the type of node used. One node type, the polynomial node, is chosen because a polynomial is a general way to represent a nonlinearity. The parameters for the polynomial node are chosen so that network based on these nodes is stable. The second type of node is based on a sigmoid function. Sigmoid functions are common in neural network studies \cite{hadaeghi2019}, although the form of our node is not the same as the most commonly used sigmoid function. The form of the nonlinearities in these two node types is different enough that our results should be general for different types of reservoir computers.

The polynomial reservoir computer is described by
\begin{equation}
\label{res_comp}
\begin{split}
& \frac{{d{r_i}\left( t \right)}}{{dt}} = \\ & \lambda \left[ {{p_1}{r_i}\left( t \right) + {p_2}r_i^2\left( t \right) + {p_3}r_i^3\left( t \right) + \sum\limits_{j = 1}^M {{A_{ij}}{r_j}\left( t \right)}  + {w_i}s\left( t \right)} \right].
\end{split}
\end{equation}
The $r_i(t)$'s  are node variables, $A$ is an adjacency matrix indicating how the nodes are connected to each other, and ${\bf W}=[w_1, w_2, ... w_M]$ is a vector that described how the input signal $s(t)$ is coupled to each node. The constant $\lambda$ is a time constant, and there are $M=100$ nodes. For the simulations described here, $p_1=-3$, $p_2=1$, $p_3=-1$ and $\lambda$ was set to minimize testing errors for different input signals. The parameters $p_1$, $p_2$ and $p_3$ were chosen to give a small value of the testing error, and they were kept the same for different input signals. The properties of $A$ will be varied to understand how the performance of the reservoir computer depends on the form of $A$.

Equation (\ref{res_comp}) was numerically integrated using a 4'th order Runge-Kutta integration routine with a time step of 0.1.  Before driving the reservoir, the mean was subtracted from the input signal $s(t)$ and the input signal  was normalized to have a standard deviation of 1.

The other reservoir computer used in this work is a map with nodes that implement a leaky tanh function. The leaky tanh node computer is described as \cite{jaeger2002}
\begin{equation}
\label{umd_comp}
\begin{split}
& {r_i}\left( {n + 1} \right) = \\ & \alpha {r_i}\left( n \right) + \left( {1 - \alpha } \right)\tanh \left( {\sum\limits_{j = 1}^M {{A_{ij}}{r_j}\left( n \right)}  + {w_i}s\left( t \right) + 1} \right).
\end{split}
\end{equation}
Again, $s(t)$ was normalized to have a mean of 0 and a standard deviation of 1. As with the polynomial nodes, the parameters were chosen to give a small value of the testing error for a fixed adjacency matrix, and we use 100 nodes.

When the reservoir computer was driven with $s(t)$, the first 2000 time steps were discarded as a transient. The next $N=10000$ time steps from each node were combined in a $N \times (M+1)$ matrix
\begin{equation}
\label{fit_mat}
{\Omega } = \left[ {\begin{array}{*{20}{c}}
{{r_1}\left( 1 \right)}& \ldots &{{r_M}\left( 1 \right)}&1\\
{{r_1}\left( 2 \right)}&{}&{{r_M}\left( 2 \right)}&1\\
 \vdots &{}& \vdots & \vdots \\
{{r_1}\left( N \right)}& \ldots &{{r_M}\left( N \right)}&1
\end{array}} \right]
\end{equation}
The last column of $\Omega $ was set to 1 to account for any constant offset in the fit. The training signal is fit by
\begin{equation}
\label{train_fit_0}
h\left( t \right) = \sum\limits_{j = 1}^M {{c_j}{r_j}\left( t \right)} 
\end{equation}
or

\begin{equation}
\label{train_fit}
{h(t)} ={\Omega } {{\bf C}}
\end{equation}
where ${h(t)} = \left[ {h\left( 1 \right),h\left( 2 \right) \ldots h\left( N \right)} \right]$ is the fit to the training signal ${g(t)} = \left[ {g\left( 1 \right),g\left( 2 \right) \ldots g\left( N \right)} \right]$ and ${{\bf C}} = \left[ {{c_1},{c_2} \ldots {c_N}} \right]$ is the coefficient vector.

The matrix ${ \Omega} $ is decomposed by a singular value decomposition 
\begin{equation}
\label{svd}
{\Omega}   = {{US}}{{{V}}^T}.
\end{equation}
where ${ U}$ is $N \times (M+1)$, ${ S}$ is  $(M+1) \times (M+1)$ with non-negative real numbers on the diagonal and zeros elsewhere, and ${ V}$ is $(M+1) \times (M+1)$.

The pseudo-inverse of ${ \Omega }$ is constructed as a Moore-Penrose pseudo-inverse \cite{penrose1955}
\begin{equation}
\label{pinv}
{ \Omega _{inv}} = {{V}}{{{S}}^{{'}}}{{U}}^T
\end{equation}
where ${{S}}^{{'}}$ is an $(M+1) \times (M+1)$ diagonal matrix constructed from ${ S}$, where the diagonal element $S^{'}_{i,i}=S_{i,i}/(S_{i,i}^2+k^2)$, where $k=1 \times 10^{-5}$ is a small number used for ridge regression \cite{tikhonov1943} to prevent overfitting. There are some guidelines for choosing $k$ \cite{golub1979}, but in this case $k$ is chosen large enough to to keep the coefficients from becoming extremely large but small enough to keep the fitting error from becoming too large.

The fit coefficient vector is then found by
\begin{equation}
\label{fit_coeff}
{{\bf C}} = {{ \Omega } _{inv}}{g(t)}
\end{equation}.

The training error may be computed from
\begin{equation}
\label{train_err}
{\Delta _{RC}} = \frac{{\left\| {  \Omega{{\bf C}} - {g(t)}} \right\|}}{{\left\| {g(t)} \right\|}}
\end{equation}
where $\left\| {} \right\|$ indicates a standard deviation. 

The training error tells us how well the reservoir computer can fit a known training signal, but it doesn't tell us anything we don't already know. To learn new information, we use the reservoir computer in the testing configuration. As an example, suppose the input signal $s(t)$ was an $x$ signal from the Lorenz system, and the training signal $g(t)$ was the corresponding $z$ signal. Fitting the Lorenz $z$ signal trains the reservoir computer to reproduce the Lorenz $z$ signal from the Lorenz $x$ signal.

We may now use as an input signal $s'(t)$ the Lorenz signal $x'$, which comes from the Lorenz system with different initial conditions. We want to get the corresponding $z'$ signal. The matrix of signals from the reservoir is now $\Omega'$. The coefficient vector ${\bf C}$ is the same vector we found in the training stage. The testing error is
\begin{equation}
\label{test_err}
{\Delta _{tx}} = \frac{{\left\| {\Omega '{\bf{C}} - z'} \right\|}}{{\left\| {z'} \right\|}}
\end{equation}
The testing error measures how accurately the reservoir computer actually solves a problem.

\section{The Input Coupling Vector ${\bf W}$}
\label{input}
The coupling vector ${\bf W}=[w_1, w_2, ... w_M]$ describes how the input signal $s(t)$ couples into each of the nodes. We want to look only at the effect of varying the coupling between nodes in the reservoir computer, so ${\bf W}$ is kept fixed. We have found that setting all the elements to +1 or -1 yields a larger reservoir computer testing error than setting the odd elements of ${\bf W}$ to +1 and the even elements of ${\bf W}$ to -1, so the second method (odd=+1, even=-1) was used. This choice was arbitrary, and other choices of ${\bf W}$ could be made. Below we show how the reservoir computer performs for our choice of input coupling vector compared to a random input vector.

\section{Characterizing the Adjacency Matrix $A$}
\label{adjacency}
 As described above, the reservoir contains 100 nodes, so the size of $A$ is $M \times M =100 \times 100$. 

 The diagonal elements of $A$ are all 0. Initially, all the off diagonal elements (network edges) of $A$ are set to +1 or 0. The initial network defined by $A$ is connected, meaning that for each pair of nodes there is a path between them. Different configurations of the network are created by flipping some of the edges between nodes  from +1 to -1. The number of elements to be flipped, $N_f$,  is chosen and then the particular elements to be flipped are chosen randomly from all the elements that have the value +1 to give many realizations of the adjacency matrix for each value of $N_f$. After the edges are flipped, the adjacency matrix is renormalized so that the absolute value of the largest real part of the matrix eigenvalues is 0.5.
 
 Different networks with the same $N_f$ will reveal a range of testing errors for a fixed value of $N_f$. For each $N_f$ value, the network is initialized to have the same adjacency matrix ${ A }$ , with all edges equal to +1 or 0. For each $N_f$, 20 different sets of $N_f$ edges from the nonzero network edges are randomly chosen to be flipped, and the testing error is calculated for each of the 20 different versions of the network.
 
 We choose to characterize the network by the fraction of the edges flipped, or $\varepsilon_f$. For some values of $\varepsilon_f$, there may be ways to permute the network nodes and their attached edges that leave the network unchanged. If this type of permutation is possible, we say the network contains symmetries, and we use the number of symmetries in place of $\varepsilon_f$ to characterize the network.

\section{Analysis Methods}
\label{analysis}
Besides calculating the training error for different reservoir computers, we analyze the reservoirs by the number of symmetries in the network, by the rank of the covariance matrix of the network and by the memory capacity of the network. We proceed to describe these different methods.

\subsection{Symmetry}
\label{symmetry}
\begin{figure}[h!]
  \includegraphics[scale=0.5]{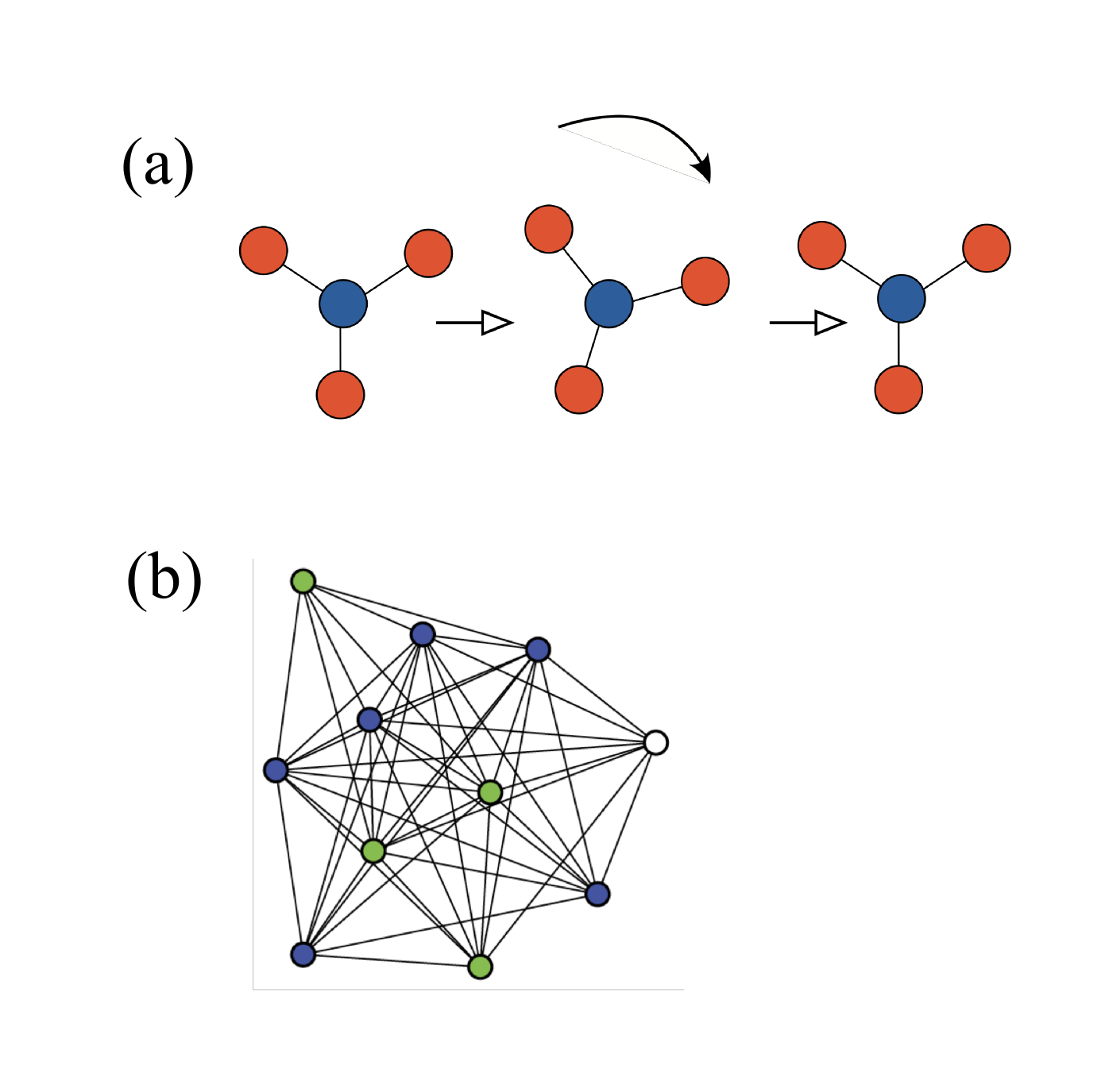}
  \caption{  (a) Simple network with rotation and mirror symmetries. (b) A more complex network where the symmetries are not obvious.
  \label{fig:1}}
\end{figure}

Symmetries in networks can have a dramatic effect on the dynamics. Here we use the concept of symmetry from graph theory \cite{GolubitskyBOOKII}, where a symmetry is a permutation of the nodes of the network along with the edges attached to the nodes which leave the network unchanged. This is shown in Fig.~\ref{fig:1}. A simple 4-node network is shown in Fig.~\ref{fig:1}(a).  The six symmetries are obvious.  Along with the identity (no permutations), there are two rotations and three mirror symmetries. 

Symmetries are easy to see with small networks, but with larger networks (7 or more nodes), the detection of symmetries is difficult and quickly becomes humanly impossible, as networks with more than 10 nodes can have millions or more symmetries (see Fig.~\ref{fig:1}(b)) . There is however an algorithm \cite{Stein} which can quickly determine the number of symmetries and give all possible permutation matrices from the matrix of connections. We use this algorithm here.

The reason symmetries can affect the dynamics of the network can be seen from the equations of motion. Let's apply a symmetry permutation $P$ to the reservoir system. If ${\bf r}=(r_1,\ldots,r_M)$, then $P{\bf r}=(r_{\pi(1)},\ldots,r_{\pi(M)})$, where $\pi(i)$ is a permutation of $(1,\ldots,M)$ into a different order. That is, $P$ moves the components of $\bf r$ around into a different ordering. Note that if $P$ is a symmetry of the network, then the network coupling matrix $A$ (think of an adjaceny matrix or Laplacian, for instance) must remain unchanged under the action of $P$, thus $PAP^T=A$, recalling that $P^{-1}=P^T$. This means $A$ and $P$ commute: $PA=AP$.

The equations of motion used in this paper are of the form (see Eqs.(\ref{res_comp}) and (\ref{umd_comp})),

\begin{equation}
\label{eq:genresveq}
\frac{d\bf r}{dt}\,=\,{\bf F}({\bf r})+A {\bf H}({\bf r})+{\bf W}s(t),
\end{equation}

\noindent where ${\bf F}({\bf r})$ is the node vector field, ${\bf H}({\bf r})$ is the coupling function, $A$ is the coupling matrix, and ${\bf W}s(t)$ is the weighted driving term. If the weights for the drive term are invariant under $P$, i.e., $P{\bf W}={\bf W}$ and we apply a symmetry $P$ to Eq.~(\ref{eq:genresveq}) and recall that $A$ and $P$ commute and the functions ${\bf F}$ and ${\bf H}$ are the same for all nodes, we get,

\begin{equation}
\label{eq:Pgenresveq}
\frac{dP\bf r}{dt}\,=\,{\bf F}(P{\bf r})+A{\bf H}(P{\bf r})+{\bf W}s(t)
\end{equation}

\noindent In other words, the permuted nodes $P{\bf r}$ have the same equation of motion as the original nodes. The consequence is that if the subsets of nodes that are permuted are started in synchronized state ($r_{\pi (i)}(t_0)=r_i(t_0)$), they will remain synchronized.  This dynamic is called {\em flow invariance}. The state where symmetry-related nodes synchronize is called {\em cluster synchronization}, where the nodes related by symmetry permutations are synchronized among themselves, but are not synchronized with nodes in other clusters \cite{GolubitskyBOOKII, PecoraClusterSyncNat2014, ChoNishikMotterPRL2017}.

If this synchronized state is stable, then it is possible that the system will evolve into it. The dimension of the network will then be lower since multiple nodes will follow identical trajectories.  Note that even if the symmetries are approximate, i.e. the components of $A$ and/or the node dynamics ${\bf F}$, and/or the weights ${\bf w}$ vary only slightly from a symmetric case there can still be approximate synchronization \cite{SorrentinoApproxSymm2016} where the trajectories of nodes related by symmetry permutations tend to closely follow an average trajectory. This still results in a reduction of dimension and complexity. 
 
 \subsection{Covariance Rank of $\Omega$}
\label{cov_rank}
We may also characterize the matrix of reservoir computer signals $\Omega$. The individual columns of $\Omega$ will be used as a basis to fit the training signal $g(t)$. The columns of $\Omega$ may be correlated with each other, so we would like to know the number of uncorrelated columns in $\Omega$. 

Principle component analysis \cite{joliffe2011} states that the eigenvectors of the covariance matrix of $\Omega$, $\Theta=\Omega^T\Omega$, form an uncorrelated basis set. The rank of the covariance matrix tells us the number of uncorrelated vectors. Therefore, we will use the rank of the covariance matrix of $\Omega$,
\begin{equation}
\label{rank}
\Gamma  = {\rm{rank}}\left( {\Omega ^T\Omega } \right)
\end{equation}
to characterize the reservoir matrix $\Omega$. We calculate the rank using the MATLAB rank() function, which returns the number of singular values above a certain threshold. The threshold is $\gamma_r={D_{\max }}\delta \left( {{\sigma _{\max }}} \right)$, where $D_{max}$ is the largest dimension of $\Omega$ and $\delta(\sigma_{max})$ is the difference between the largest singular value of $\Omega$ and the next largest double precision number. 

\subsection{Memory}
\label{mem_cal}
Memory capacity, as defined in \cite{jaeger2002}, is considered to be an important quantity in reservoir computers. Memory capacity is a measure of how well the reservoir can reproduce previous values of the input signal.

The memory capacity as a function of delay is
\begin{equation}
\label{memdel}
{\rm{M}}{{\rm{C}}_k} = \frac{{\sum\limits_{n = 1}^N {\left[ {s\left( {n - k} \right) - \overline s } \right]\left[ {{g_k}\left( n \right) - \overline {{g_k}} } \right]} }}{{\sum\limits_{n = 1}^N {\left[ {s\left( {n - k} \right) - \overline s } \right]\sum\limits_{n = 1}^N {\left[ {{g_k}\left( n \right) - \overline {{g_k}} } \right]} } }}
\end{equation}
where the overbar indicator indicates the mean. The signal $g_k(n)$ is the fit of the reservoir signals $r_i(n)$ to the delayed input signal $s(n-k)$. The memory capacity is
\begin{equation}
\label{memcap}
{\rm{MC}} = \sum\limits_{k = 1}^\infty  {{\rm{M}}{{\rm{C}}_k}} 
\end{equation}

Input signals such as the Lorenz $x$ signal contain correlations in time, which will cause errors in the memory calculation, so in eq. (\ref{memdel}), $s(n)$ is a random signal uniformly distributed between -1 and +1. The node parameters were optimized to minimize the testing error for fitting the reservoir signals to $s(n-1)$. There are some drawbacks to defining memory in this way; the reservoir is nonlinear, so its response will be different for different input signals, and the node parameters for the memory calculation are not the same as for calculations with other input signals. Nevertheless, this memory definition is the standard definition used in the field of reservoir computing.

\section{Input signals}
\label{signals}
The first system we used to generate input and training signals is the Lorenz system \cite{lorenz1963}
\begin{equation}
\label{lorenz}
\begin{array}{l}
\frac{{dx}}{{dt}} = {c_1}y - {c_1}x\\
\frac{{dy}}{{dt}} = x\left( {{c_2} - z} \right) - y\\
\frac{{dz}}{{dt}} = xy - {c_3}z
\end{array}
\end{equation}

with $c_1$=10, $c_2$=28, and $c_3$=8/3. The equations were numerically integrated with a time step of $t_s=0.02$.

The second system is a nonlinear map acting on a random signal, taken from \cite{rodan2011}
\begin{equation}
\label{narma}
\begin{array}{*{20}{l}}
{x\left( k \right) = {\rm{random}}\left[ {0,0.5} \right]}\\
{y\left( {k + 1} \right) = 0.3y\left( k \right) + 0.05{y^2}\left( k \right) + 1.5{x^2}\left( k \right) + 0.1}
\end{array}
\end{equation}
This system is commonly used as a test of the ability of a reservoir computer to fit a signal.

\section{Simulations: Flipping Network Edges}
\label{simulations}
Because the reservoir is nonlinear, changing the adjacency matrix can have a complicated effect on the testing error. In order to get good statistics, each time the number of network edges $N_f$ to be flipped was chosen, 20 different adjacency matrices were generated with with $N_f$ randomly flipped edges. The graphs below show all 20 values of the testing error for each number of edges flipped.

We begin with a $100 \times 100$ adjacency matrix with 9800 of the network edges equal to +1. The large number of nonzero edges made it more likely that the network would have symmetries. Once the number of nonzero edges was chosen, the specific nonzero edges were chosen randomly until a network with a large number of symmetries was found.  All of the diagonal elements are 0.

 If the number of nonzero network edges is $N_1$, then the fraction of edges flipped is $\varepsilon_f=N_f/N_1$. For some values of $\varepsilon_f$, the network contains symmetries (see section \ref{symmetry}). For networks that contain symmetries, the network will be characterized by the number of symmetries, $\zeta_s$. When no edges were flipped, the network contained $9.2678 \times 10^{51}$ symmetries, calculated using the methods from \cite{Stein}. For networks that contain only one symmetry, the identity, the network will be characterized by the fraction of elements flipped, $\varepsilon_f$.

\subsection{Comparison to other networks}
\label{net_comp}
The network we choose appears to be specialized, so we must ask if the results will apply to other reservoir computers. As an alternative to our configuration, we simulated reservoir computers where
\begin{enumerate}
 \item 98\% of the edges were $\pm 1$ and the elements of the input vector ${\bf W}$ alternated between +1 and -1; 
 \item 98\% of the edges were $\pm 1$ and the elements of the ${\bf W}$ were all +1; 
 \item 98\% of the edges were $\pm 1$ and the elements of the ${\bf W}$ were chosen from a random uniform distribution between $\pm 1$; and 
 \item 20\% of the network edges were nonzero, chosen from a uniform distribution between $\pm 1$ and the elements of the ${\bf W}$ were chosen from a random uniform distribution between $\pm 1$. 
 \end{enumerate}
 The last choice is typical of reservoir computer simulations that have been published. 

\begin{figure}[ht]
\centering
\includegraphics[scale=0.8]{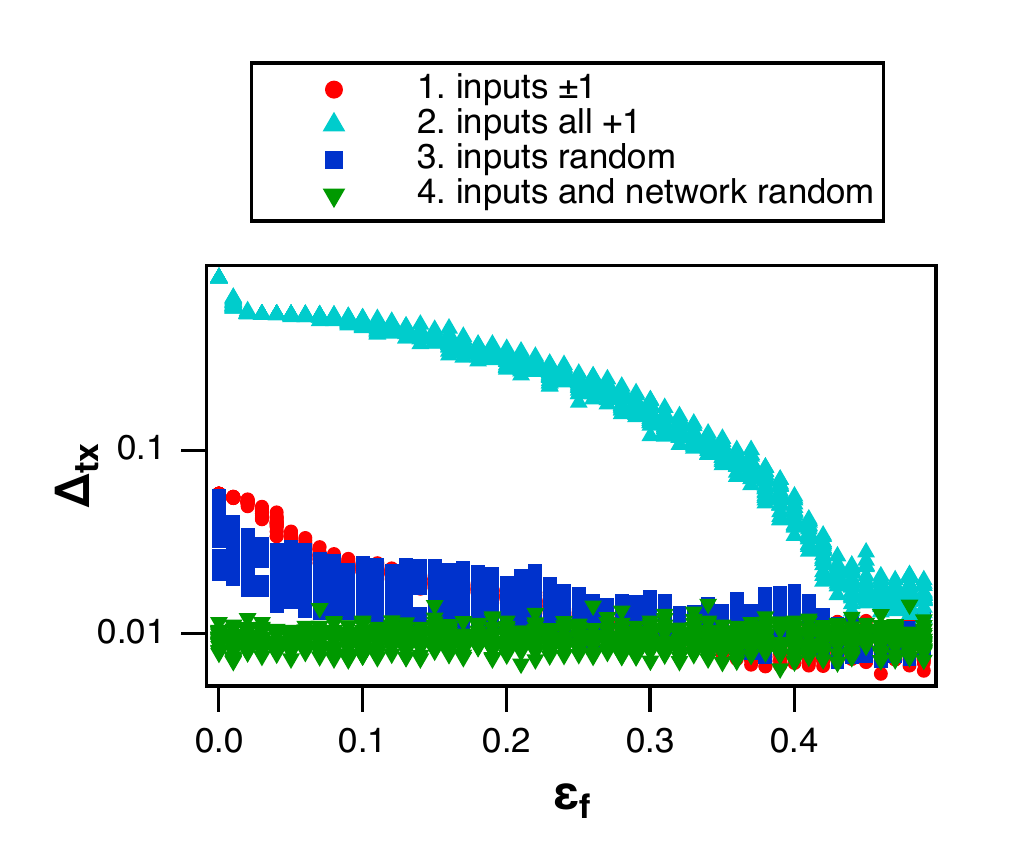} 
  \caption{ \label{varyw} Testing error $\Delta_{tx}$ as a function of fraction of network edges flipped $\varepsilon_f$ from +1 to -1 for several different types of input vector ${\bf W}$. For the inputs and network random (upside down green triangles, number 4 in the legend), the $\varepsilon_f$ scale does not apply. The random input and network data was just plotted on the same axes for comparison.
}
  \end{figure}
Figure \ref{varyw} shows the testing error $\Delta_{tx}$ as a function of fraction of edges flipped $\varepsilon_f$ for the first three cases. The fourth situation, where the input vector and network were random, is also plotted on the same axes for comparison. When all the elements of the input vector are +1, the testing error is much larger than when the elements are $\pm 1$. Choosing elements of the input vector from a random distribution gives a smaller testing error when $\varepsilon_f$ is small, but for larger values of $\varepsilon_f$ the testing errors are in the same range as when all elements are $\pm 1$. When the network edges and input vector elements are all chosen from a random distribution, the testing error is in the same range as when $\varepsilon_f=0.5$. 

Figure \ref{varyw} does show that the standard practice of choosing all network edges randomly does give the best result, but when the fraction of edges flipped is 50\%, our networks of all $\pm 1$ work just as well, so while networks need some randomness, they do not have to be completely random. 

Figure \ref{varyw} indicates that just using a completely random network is a good choice, but there may  be restrictions on choosing network edges in experimental situations, so it is important to know how to create good networks that are not random. Finally, without a theory, the only way to know if a network choice is optimum is to simulate different networks and measure their performance.

\subsection{Testing error vs. Number of Symmetries}
\label{sym_sim}
\subsubsection{Polynomial Nodes}

\begin{figure}[ht]
\centering
\includegraphics[scale=0.4]{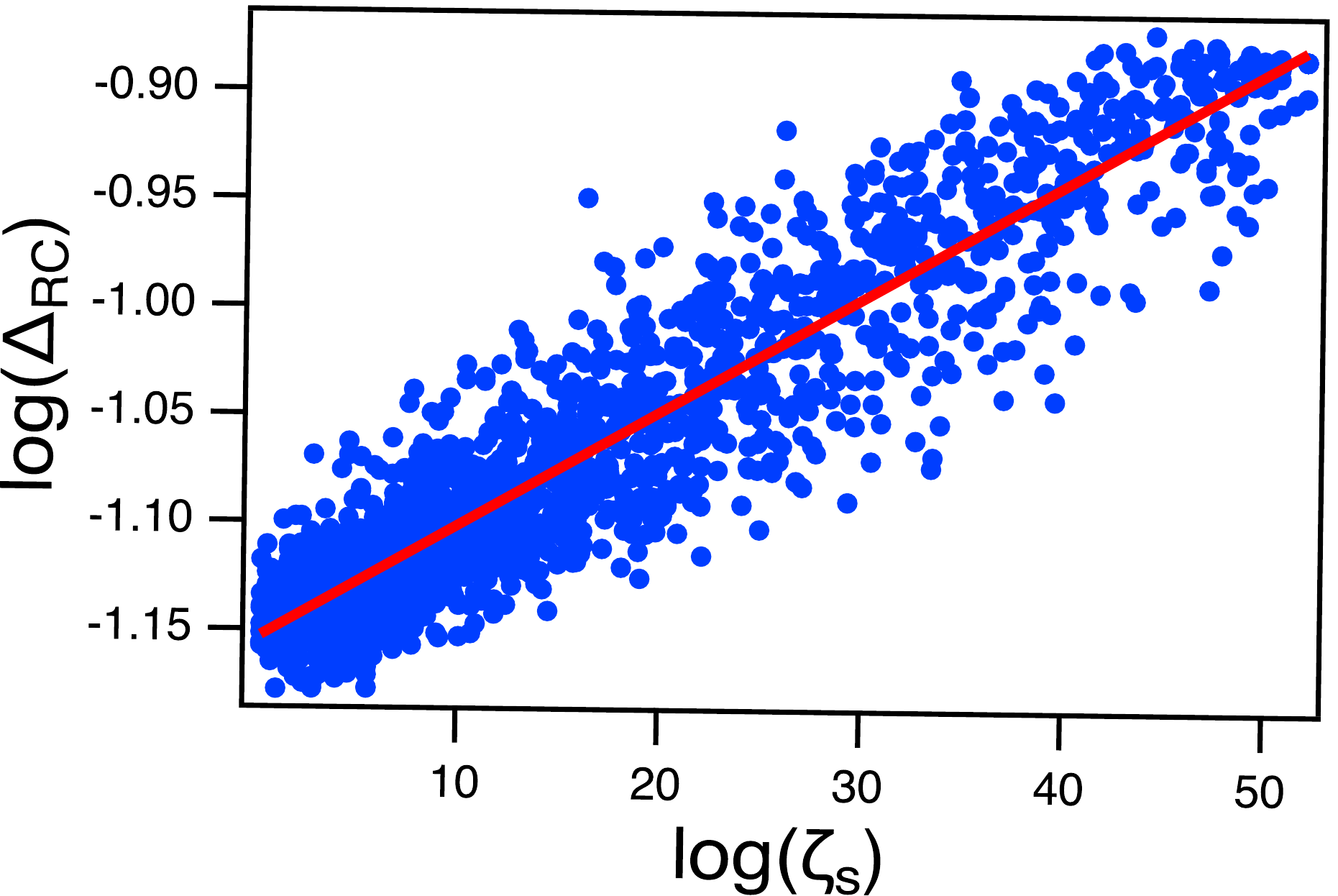} 
  \caption{ \label{symm_lor_nleq}
Reservoir computer testing error $\Delta_{tx}$ vs. number of symmetries $\zeta_s$ in the network for a reservoir computer with polynomial nodes when input signal $s(t)$ is the Lorenz $x$ signal and the training signal $g(t)$ is the Lorenz $z$ signal. The logarithms are base 10. }
  \end{figure}

These simulations used a reservoir computer with the polynomial nodes (eq. \ref{res_comp}). The value of $\lambda$ was chosen to minimize testing error by matching the time scale of the reservoir to the time scale of the input signal. Figure \ref{symm_lor_nleq} shows the log base 10 of testing error $\Delta_{tx}$ as a function of the log base 10 of the number of symmetries $\zeta_s$ for a reservoir computer with polynomial nodes when the input signal $s(t)$ is the Lorenz $x$ signal and the training signal $g(t)$ is the Lorenz $z$ signal. For this combination of input and training signals, $\lambda=1.4$. 

\begin{figure}[ht]
\centering
\includegraphics[scale=0.4]{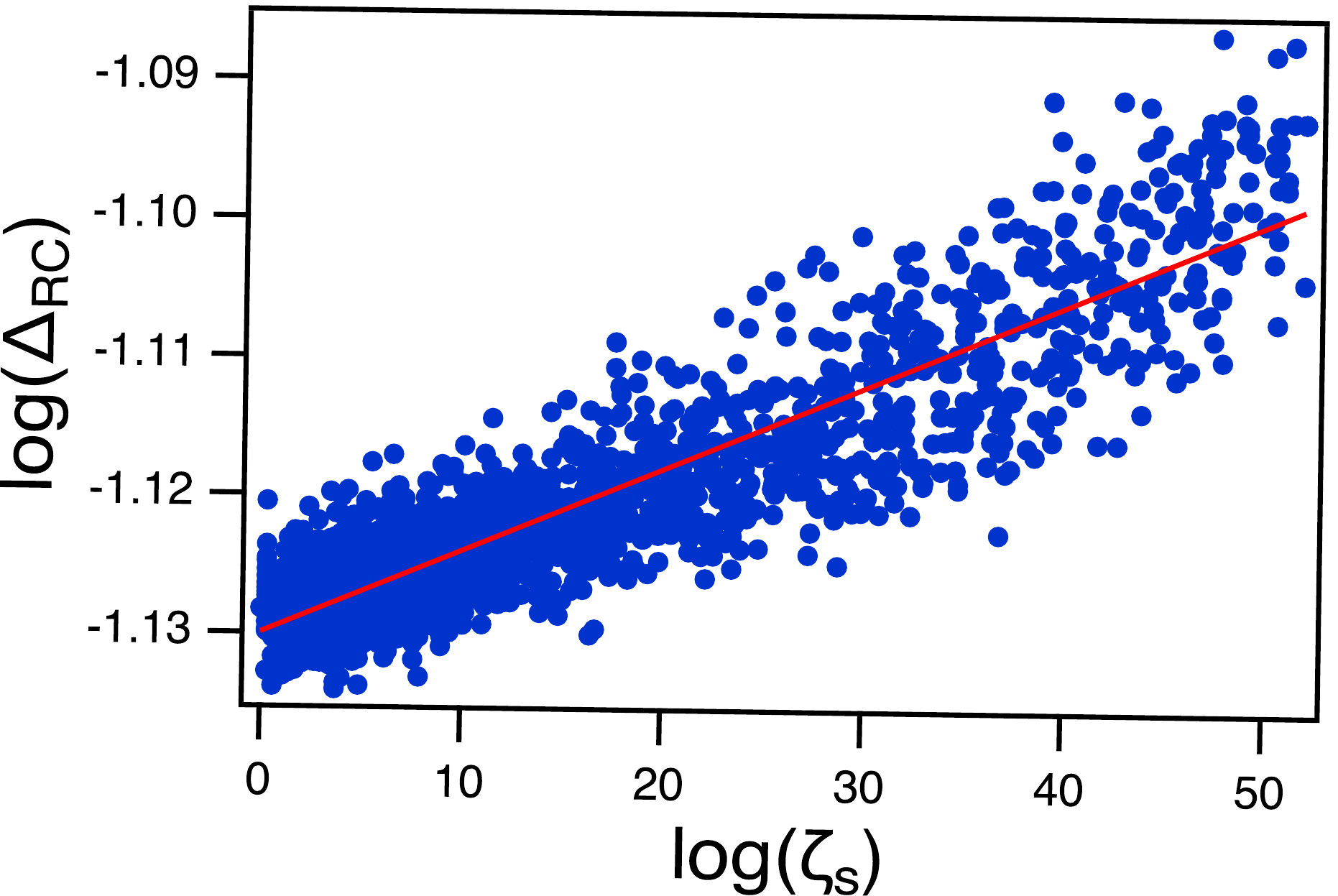} 
  \caption{ \label{symm_narma_nleq}
Reservoir computer testing error $\Delta_{tx}$ vs. number of symmetries $\zeta_s$ in the network for a reservoir computer with polynomial nodes when input signal $s(t)$ is the random signal $x(k)$ from eq. (\ref{narma}) and the training signal $g(t)$ is the $y(k)$ signal from this equation. The logarithms are base 10. }
  \end{figure}

Figure \ref{symm_narma_nleq} shows the testing error for the reservoir computer with polynomial nodes when the input signal $s(t)$ is the nonlinear map signal $x(k)$ (eq. \ref{narma}) and the training signal $g(t)$ is the nonlinear map signal $y(k)$. The value of $\lambda$ was set to 5 to minimize the testing errors.
  
Comparing figures \ref{symm_lor_nleq} and \ref{symm_narma_nleq}, in both cases the testing error $\Delta_{tx}$ increases as the number of symmetries $\zeta_s$ increases, so that when the are more ways that the individual nodes can be re-arranged without changing the network structure, the testing error is larger. 

\subsubsection{Leaky Tanh Nodes}

The reservoir computer with leaky tanh nodes (eq. \ref{umd_comp}) follows the same trends as the reservoir computer with polynomial nodes, but some of the details are different. When the input signal was the Lorenz $x$ signal, the parameter $\alpha$ in eq. (\ref{umd_comp}) was set to 0.35 and the spectral radius was set to 1.0. These parameters were found by minimizing the testing error.

\begin{figure}[ht]
\centering
\includegraphics[scale=0.8]{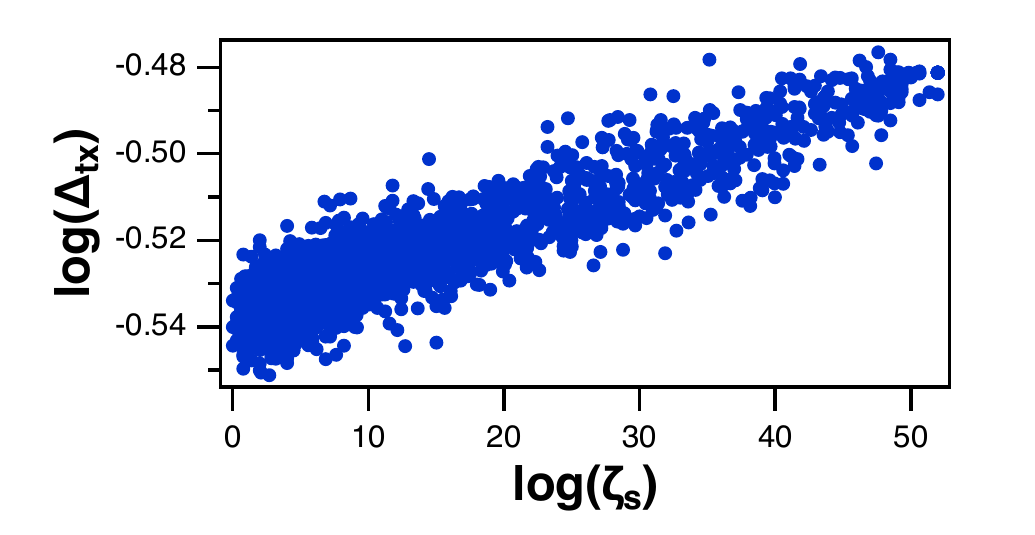} 
  \caption{ \label{symm_lor_umd}
Reservoir computer testing error $\Delta_{tx}$ vs. number of symmetries $\zeta_s$ in the network for a reservoir computer with leaky tanh nodes when input signal $s(t)$ is the Lorenz $x$ signal and the training signal $g(t)$ is the Lorenz $z$ signal.  The logarithms are base 10.}
  \end{figure}
  
\begin{figure}[ht]
\centering
\includegraphics[scale=0.8]{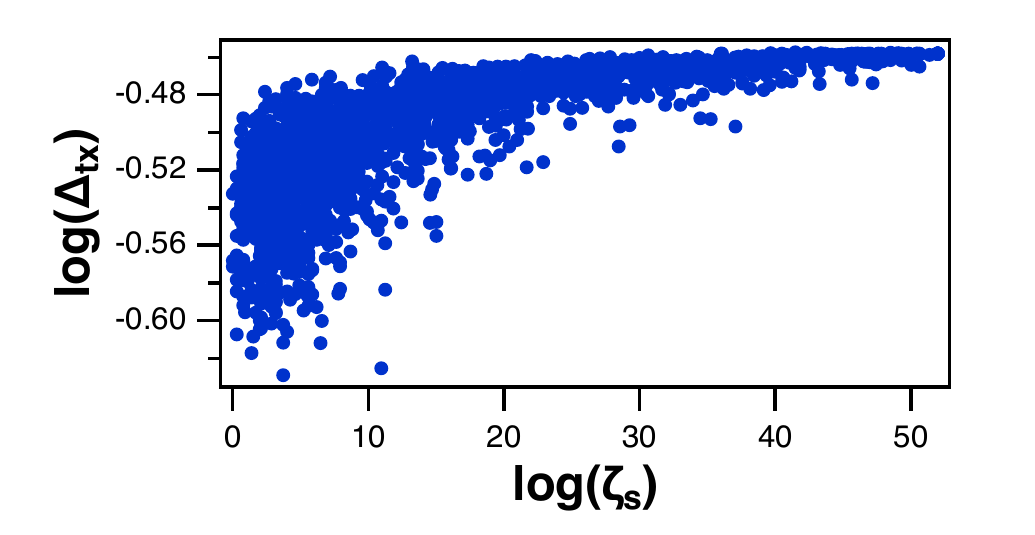} 
  \caption{ \label{symm_narma_umd}
Reservoir computer testing error $\Delta_{tx}$ vs. number of symmetries $\zeta_s$ in the network for a reservoir computer with leaky tanh nodes when input signal $s(t)$ is the random input signal  $x(k)$ from eq. (\ref{narma}) and the training signal $g(t)$ is the $y(k)$ output signal. The logarithms are base 10. }
  \end{figure}

 Figure \ref{symm_narma_umd} shows the testing error when the input signal is the nonlinear map signal $x(k)$ (eq. \ref{narma}) and the training signal $g(t)$ is the nonlinear map signal $y(k)$. In this case, the log-log plot of testing error vs number of symmetries is not linear, but we do not know why the plot is nonlinear.
 
\subsection{Testing error vs. Fraction of Edges Flipped}
\label{flip}
If more than 100 edges were flipped in the network used in the previous section, we could find only one symmetry, the identity. For larger numbers of flipped edges, we plot the testing error $\Delta_{tx}$ as a function of the fraction of edges flipped from +1 to -1, $\varepsilon_f$. 

\subsubsection{Polynomial and Linear Nodes}
In this section we consider two types of node. We study the polynomial nodes of eq. (\ref{res_comp}) and a reservior computer with linear nodes. The linear nodes are also described by eq. (\ref{res_comp}), but with parameters $p_1=-3$, $p_2=0$ and $p_3=0$. The difference in the testing error $\Delta_{tx}$ will reveal how the reservoir computer performance depends on nonlinearity.

\begin{figure}[ht]
\centering
\includegraphics[scale=0.8]{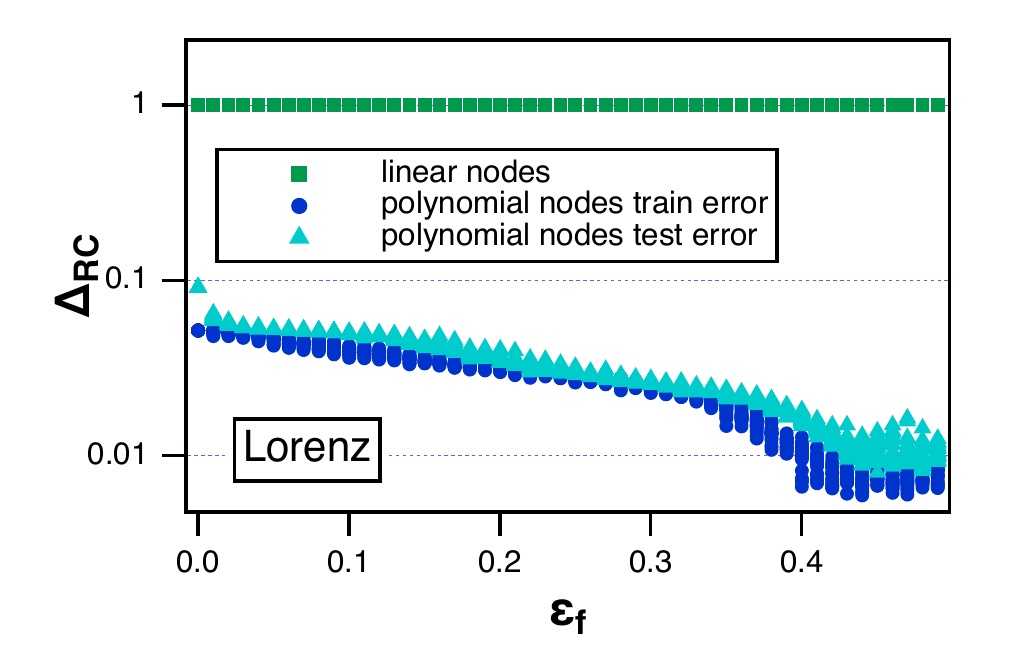} 
  \caption{ \label{lor_nleq_100_err}
Blue circles are the reservoir computer training  error $\Delta_{tx}$ vs. fraction of edges $\varepsilon_f$ in the network flipped from +1 to -1 for a reservoir computer with polynomial nodes when input signal $s(t)$ is the Lorenz $x$ signal and the training signal $g(t)$ is the Lorenz $z$ signal. The blue dots indicate the testing error $\Delta_{tx}$. The green squares are for the same system when the reservoir computer nodes are linear, that is in eq. (\ref{res_comp}) $p_1=-3$, $p_2=0$ and $p_3=0$.}
  \end{figure}

 Figure \ref{lor_nleq_100_err} shows the testing error $\Delta_{tx}$ vs. the fraction of edges flipped $\varepsilon_f$ for a reservoir computer when the nodes were described by the polynomial of eq. (\ref{res_comp}). The input signal $s(t)$ was the Lorenz $x$ signal, while the training signal $g(t)$ was the Lorenz $z$ signal. The figure also shows the testing error when the nodes were linear, that is $p_1=-3$, $p_2=0$ and $p_3=0$. 
  
  As a larger fraction of edges in the network are flipped from +1 to -1, the testing error plotted in figure \ref{lor_nleq_100_err} decreases if the nodes are polynomial nodes. If the nodes were linear, figure \ref{lor_nleq_100_err} shows that the testing error $\Delta_{tx}$ did not decrease. Nonlinearity is necessary for the reservoir computer to fit the Lorenz $z$ signal when the input signal was the Lorenz $x$ signal.
 
\begin{figure}[ht]
\centering
\includegraphics[scale=0.8]{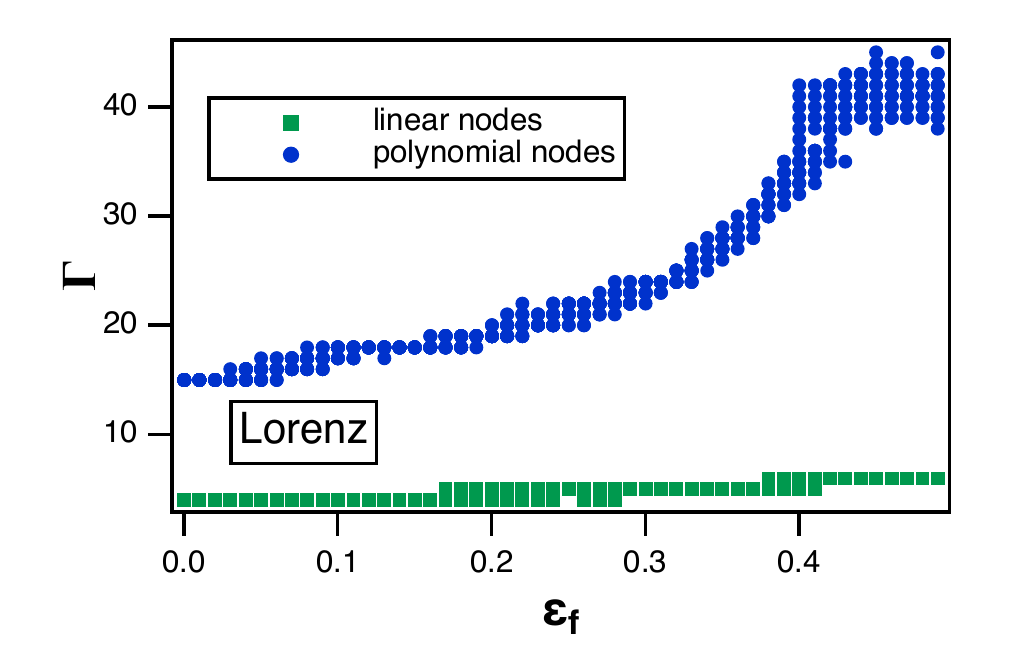} 
  \caption{ \label{nleq_rank_fraction}
Blue circles are the rank $\Gamma$ of the covariance of the reservoir computer matrix $\Omega$ vs the fraction of edges flipped $\varepsilon_f$ , when the nodes are polynomial nodes. The input signal $s(t)$ was the Lorenz $x$ variable, while the training signal $g(t)$ was the Lorenz $z$ signal. The green squares are the covariance rank when the nodes were linear.}
  \end{figure}

A possible explanation for why flipping more edges in the network from +1 to -1 reduces the testing error $\Delta_{tx}$ is shown in figure \ref{nleq_rank_fraction}, which shows the covariance rank $\Gamma$ (defined in eq. \ref{rank}) of the reservoir variables ${\bf R}(t)$ as a function of fraction of edges flipped, $\varepsilon_f$.

 When the nodes are polynomial nodes, figure \ref{nleq_rank_fraction} shows that the rank $\Gamma$ of the covariance of the reservoir matrix $\Omega$ increases with the fraction of edges flipped, while the rank when the nodes are linear increases only slightly, with a maximum rank of 6. The linear reservoir variables span a much lower dimension than the polynomial reservoir variables, which may be why the polynomial reservoir does a better job of fitting the Lorenz $z(t)$ signal in figure \ref{lor_nleq_100_err}.

\begin{figure}[ht]
\centering
\includegraphics[scale=0.8]{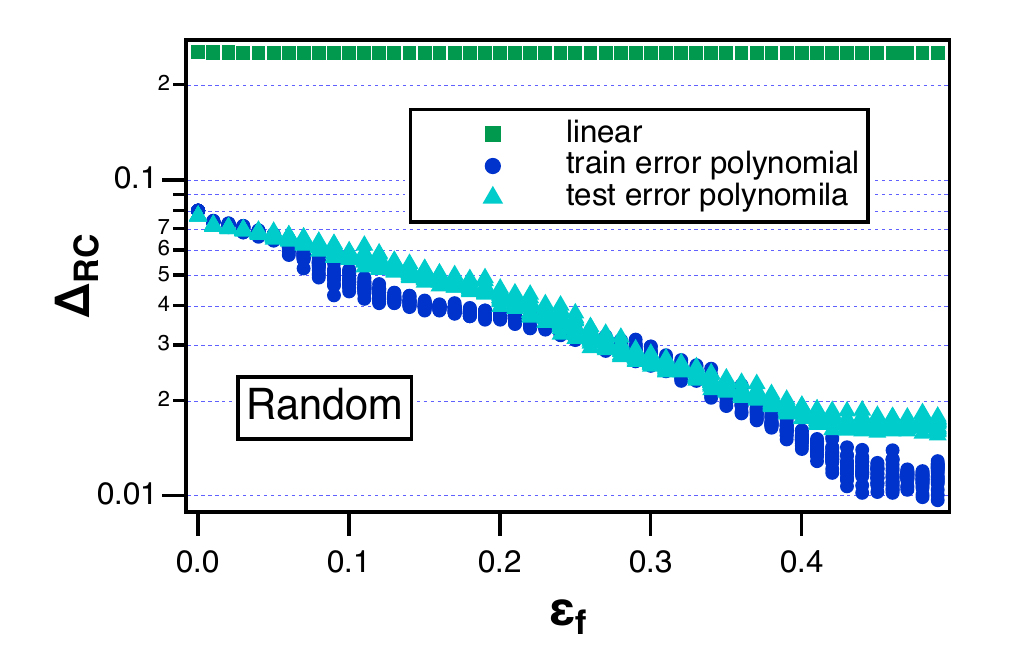} 
  \caption{ \label{narma1_nleq_100_err}
Blue circles are the reservoir computer testing error $\Delta_{tx}$ vs. fraction of edges $\varepsilon_f$ in the network flipped from +1 to -1 for a reservoir computer with polynomial nodes when input signal $s(t)$ was the random $x(k)$ signal from eq. (\ref{narma})  and the training signal is the $y(k)$ signal. The light blue triangles indicate the testing error $\Delta_{tx}$. The green squares are for the same system when the reservoir computer nodes are linear, that is in eq. (\ref{res_comp}) $p_1=-3$, $p_2=0$ and $p_3=0$.}
  \end{figure}

When the input signal $s(t)$ for the reservoir comes from the random $x(k)$ signal from eq. (\ref{narma})  and the training signal is the $y(k)$ signal, figure \ref{narma1_nleq_100_err} shows the testing error $\Delta_{tx}$ as a function of fraction of edges $\varepsilon_f$  in the network flipped from +1 to -1, for both polynomial nodes and linear nodes.
 
\begin{figure}[ht]
\centering
\includegraphics[scale=0.4]{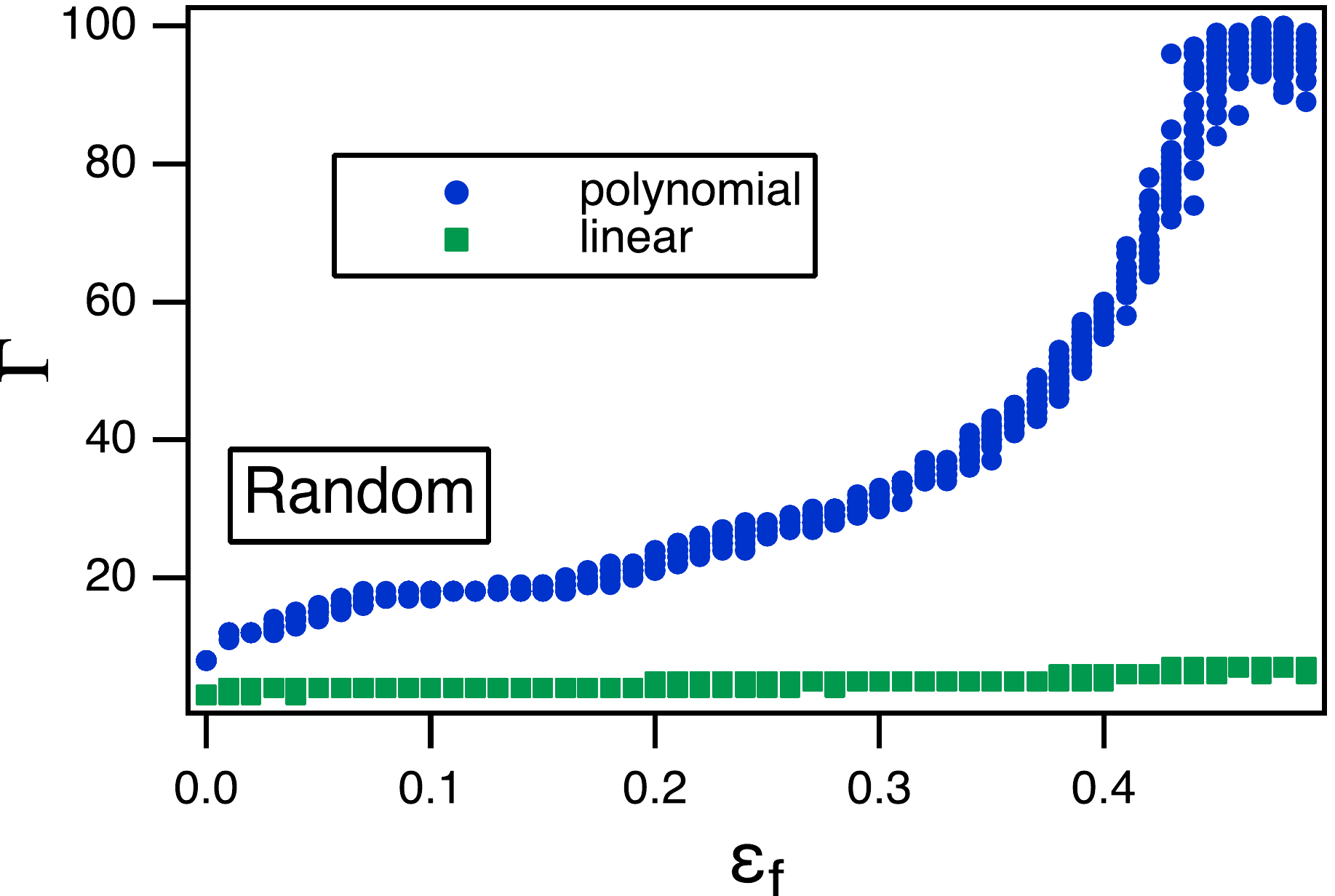} 
  \caption{ \label{narma1_rank_fraction}
Blue circles are the covariance rank $\Gamma$ of the  matrix of the reservoir computer variables $\Omega$ as defined in eq. (\ref{rank}), when the nodes are polynomial nodes. The input signal $s(t)$ was the random $x(k)$ signal from eq. (\ref{narma})  and the training signal is the $y(k)$ signal. The green squares are the covariance rank when the nodes were linear. The fraction of edges flipped from +1 to -1 in the network is $\varepsilon_f$.}
  \end{figure}

Figure \ref{narma1_rank_fraction} shows the covariance rank $\Gamma$ as a function of fraction of edges flipped, $\varepsilon_f$ when the reservoir input signal is the random $x(k)$ signal from eq. (\ref{narma})  and the training signal is the $y(k)$ signal. 
  
  Similar to when the polynomial nodes were driven by the Lorenz system, the testing error $\Delta_{tx}$ decreases as the fraction of edges flipped $\varepsilon_f$ increases for polynomial nodes, but not for linear nodes for the nonlinear map system. Figure \ref{narma1_rank_fraction} shows that once again, the covariance rank $\Gamma$ increases with the fraction of edges flipped for polynomial nodes, but increases only slightly for the linear nodes. The rank actually saturates at 100 for the polynomial nodes. The signal $x(k)$  is a random signal, so it makes sense that the reservoir ${\bf R}(t)$ would have a higher covariance rank when driven with the infinite dimensional random signal than when driven by the finite dimensional Lorenz signal.

\subsubsection{Leaky Tanh Nodes}

\begin{figure}[ht]
\centering
\includegraphics[scale=0.8]{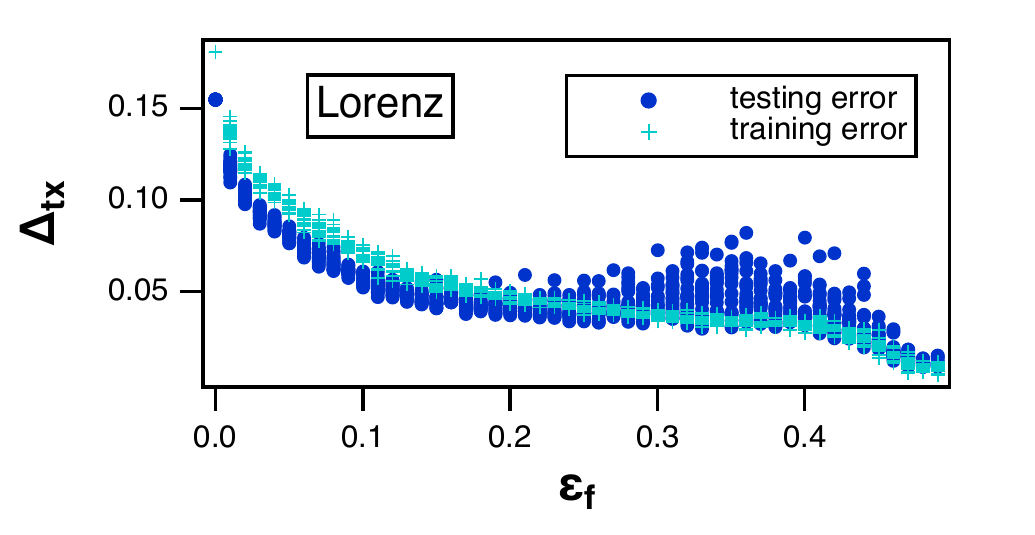} 
  \caption{ \label{lor_umd_err}
Reservoir computer testing error $\Delta_{tx}$ vs. fraction of edges $\varepsilon_f$ in the adjacency matrix $A$ flipped from +1 to -1 for a reservoir computer with leaky tanh nodes when input signal $s(t)$ is the Lorenz $x$ signal and the training signal $g(t)$ is the Lorenz $z$ signal. Blue dots are the testing error $\Delta_{tx}$, while the light blue crosses are the training error $\Delta_{RC}$.}
  \end{figure}

The testing error $\Delta_{tx}$ as a function of fraction of edges flipped $\varepsilon_f$ when the input signal $s(t)$ for the reservoir with leaky tanh nodes was the Lorenz $x$ signal and the training signal $g(t)$ was the Lorenz $z$ signal is shown in figure \ref{lor_umd_err}.

The reservoir computer testing error in figure \ref{lor_umd_err} shows a decreasing trend as the fraction of edges flipped increases, but the testing error increases between $\varepsilon_f=0.3$ to 0.4. The training error, also shown in figure \ref{lor_umd_err}, does not increase over this range. One possible reason for the larger testing error is that the reservoir may be less stable over this region of $\varepsilon_f$, so it may be more sensitive to differences in the input signal. 

\begin{figure}[ht]
\centering
\includegraphics[scale=0.8]{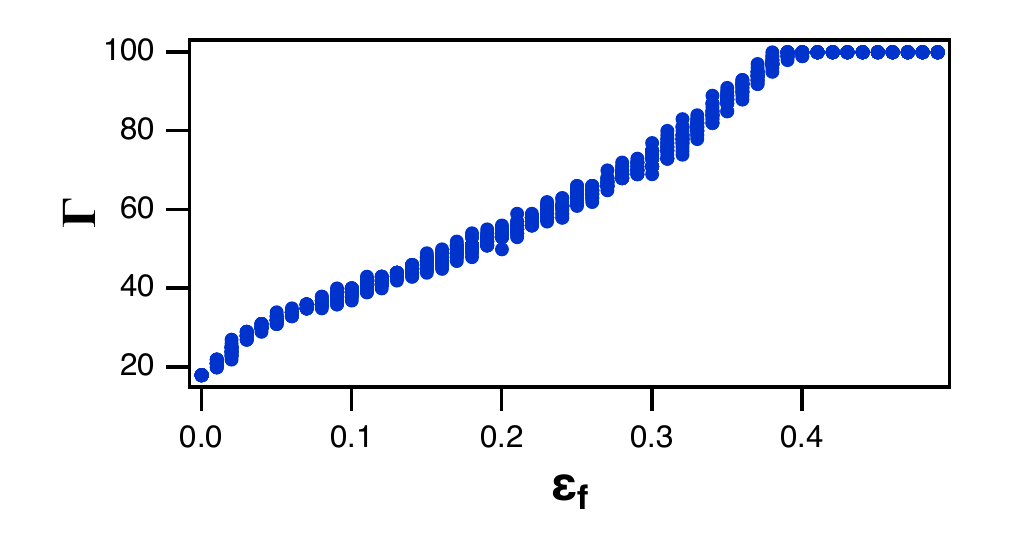} 
  \caption{ \label{lor_umd_rank_fraction}
Rank $\Gamma$ of the covariance of the matrix of the reservoir computer variables $\Omega$ as defined in eq. (\ref{rank}) as a function of the fraction of edges flipped $\varepsilon_f$ in the network, when the nodes are leaky tanh nodes. The input signal $s(t)$ was the Lorenz $x$ signal, while the training signal $g(t)$ was the Lorenz $z$ signal.   
}
  \end{figure}

Figure \ref{lor_umd_rank_fraction} shows the covariance rank $\Gamma$ as a function of fraction of edges flipped. The covariance rank saturates at its highest possible value for $\varepsilon_f > 0.4$.

 Comparing figure \ref{lor_nleq_100_err} to figure \ref{lor_umd_err}, the testing error for the reservoir using leaky tanh nodes is higher than the testing error using polynomial nodes (when driven by the Lorenz system), even though the covariance rank for the leaky tanh nodes (figure \ref{lor_umd_rank_fraction}) is higher than for the polynomial nodes (figure \ref{nleq_rank_fraction}). Clearly a higher rank is related to a lower testing error only if the node type stays the same.
  
\begin{figure}[ht]
\centering
\includegraphics[scale=0.8]{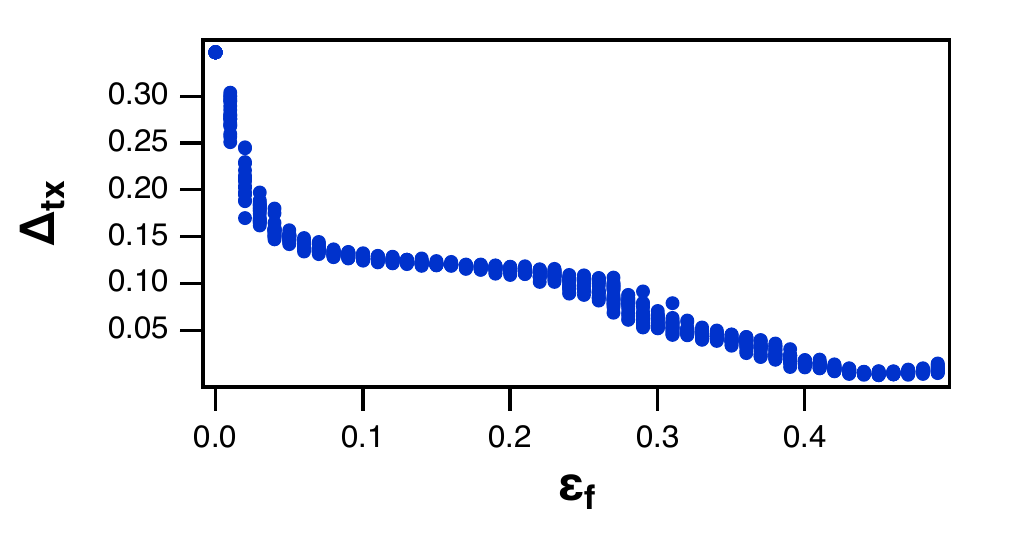} 
  \caption{ \label{narma1_umd_err}
Reservoir computer testing error $\Delta_{tx}$ vs. fraction of network edges $\varepsilon$  flipped from +1 to -1 for a reservoir computer with leaky tanh nodes when input signal $s(t)$ was the random $x$ signal from eq. (\ref{narma}), while the training signal $g(t)$ was $y(k)$ signal. }
  \end{figure}

 Figure \ref{narma1_umd_err} shows the testing error $\Delta_{tx}$ vs. the fraction of network edges $\varepsilon_f$  that have been flipped from +1 to -1 when the reservoir computer with leaky tanh nodes is driven by the random $x(k)$ signal from eq. (\ref{narma}).

    \begin{figure}[ht]
\centering
\includegraphics[scale=0.8]{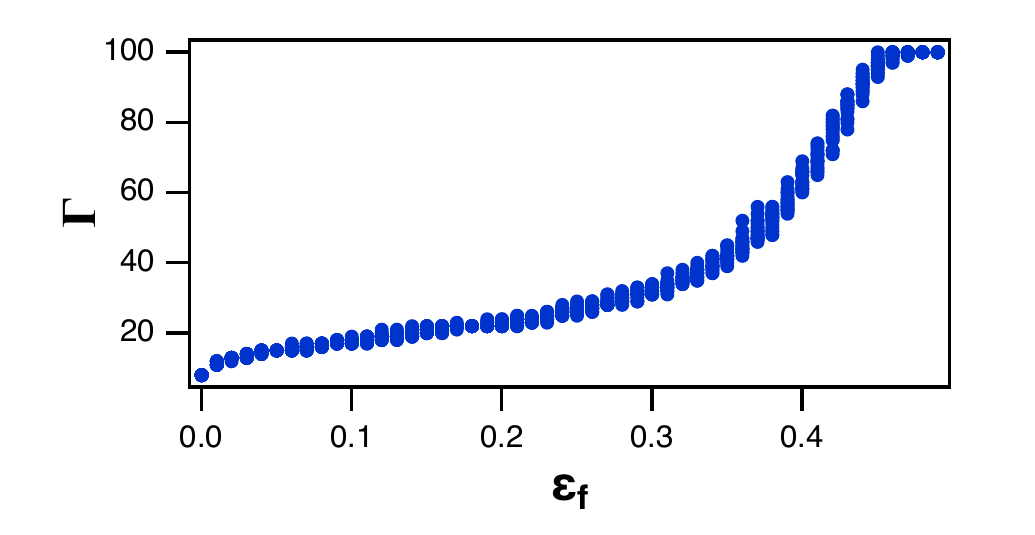} 
  \caption{ \label{narma1_umd_rank_fraction}
The rank $\Gamma$ of the covariance of the matrix of the reservoir computer variables $\Omega$ as defined in eq. (\ref{rank}), when the nodes are leaky tanh nodes, as a function of fraction of edges $\varepsilon_f$ . The input signal $s(t)$ was the random $x(k)$ signal from eq. (\ref{narma}), while the training signal $g(t)$ was the  $y(k)$ signal. }
  \end{figure}
 
As with the Lorenz input signal, when the input signal comes from the random system, the reservoir covariance rank is higher for the leaky tanh nodes but the testing error is also higher. Higher covariance rank leads to lower error when a particular input signal is used with a particular node type, but the same is not true for different combinations of node and input signal.

\subsubsection{Arbitrary Cutoff for Rank Calculation}
In the previous sections, the rank of the covariance matrix was calculated using the MATLAB rank function. This function calculates the rank as the number of singular values above a tolerance of $\gamma_r={D_{\max }}\delta \left( {{\sigma _{\max }}} \right)$, where $D_{max}$ is the largest dimension of $\Omega$ and $\delta(\sigma_{max})$ is the difference between the largest singular value of $\Omega$ and the next largest double precision number. For the covariance matrices in the previous sections, $\gamma_r=1.42 \times 10^{-12}$ for all parameters. 

\begin{figure}[ht]
\centering
\includegraphics[scale=0.7]{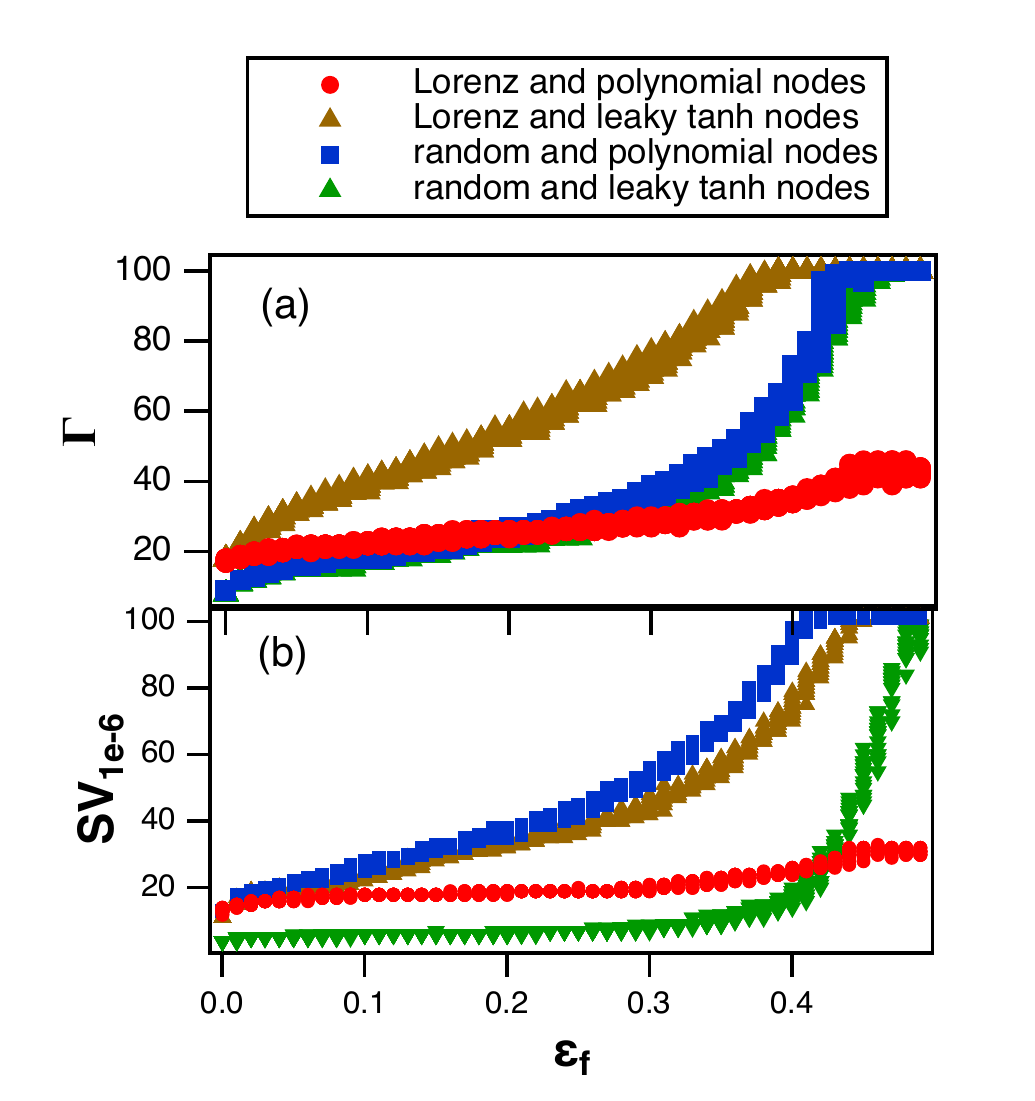} 
  \caption{ \label{ranktol}
(a) Rank $\Gamma$ of the covariance matrices from the previous four examples, calculated from the MATLAB rank function. The MATLAB function computed the rank as the number of singular values above a tolerance, which was $\gamma_r=1.42 \times 10^{-12}$ for these examples.  (b) Number of singular values ${\rm SV}_{1e-6}$ above the arbitrary threshold of $1 \times 10^{-6}$ for the same covariance matrices.}
  \end{figure}

We may see how robust these rank results are by setting a different tolerance. Figure \ref{ranktol}(a) replots the covariance ranks from the previous figures using the MATLAB tolerance of $\gamma_r=1.42 \times 10^{-12}$, while figure \ref{ranktol}(b) shows the number of singular values for the covariance matrix above the arbitrary threshold of $1 \times 10^{-6}$ times the largest singular value. This number is designated as ${\rm SV}_{1e-6}$.

The ordering of the different curves in figure \ref{ranktol} is not the same for parts (a) and (b), but all curves do increase as the fraction of edges flipped, $\varepsilon_f$, increases. 

 \subsection{Memory Capacity}
 \label{mem_cap}
\begin{figure}
\centering
\includegraphics[scale=0.8]{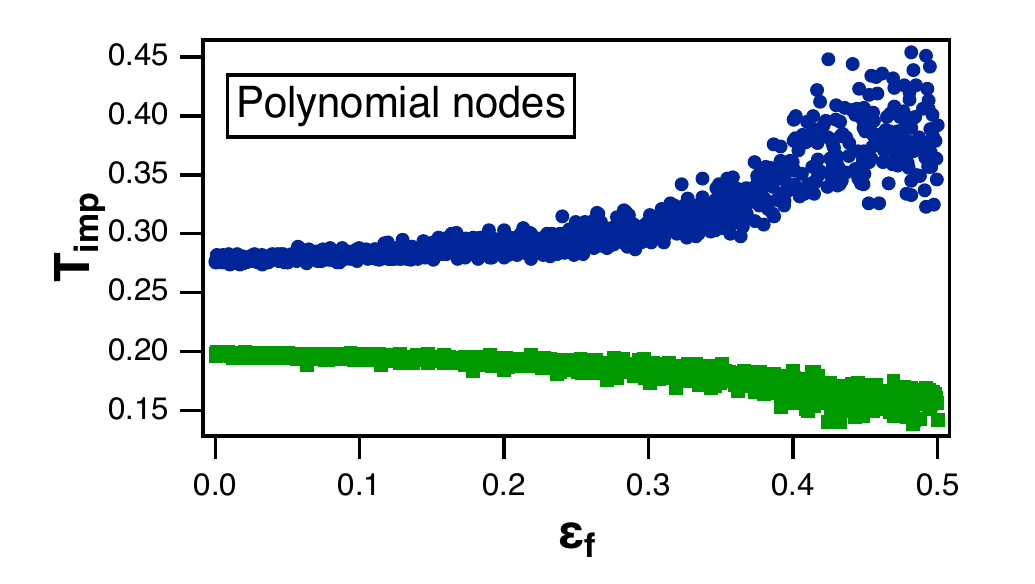} 
\caption{ \label{nleq_mem} Memory capacity MC as a function of fraction of edges flipped $\varepsilon_f$ for both node types. The time constant $\lambda$ for the polynomial nodes in eq. (\ref{res_comp}) was 6 for this simulation, while the spectral radius was 0.5. For the leaky tanh nodes, $\alpha=0.66$ and the spectral radius was 1.36.}
\end{figure}

 Figure \ref{nleq_mem} shows that the memory capacity , as defined by eq (\ref{memcap}),  increases as the fraction of edges flipped increases. Theory has suggested that longer memory leads to improved computational accuracy in a reservoir computer \cite{jaeger2002,marzen2017}. For individual node types, higher memory capacity does correlate with lower testing error, but figure \ref{nleq_mem} shows that the memory capacity for the leaky tanh nodes was higher than for the polynomial nodes, but the leaky tanh nodes gave larger testing errors. It is interesting to note that the rank for all four combinations of input signal and node type stops increasing above $\varepsilon_f > 0.4$, and the memory capacity also levels off above this value.

 \section{Varying Sparsity and Fraction Flipped}
 \label{sparse_var}
 A typical assumption for reservoir computers is that a sparse adjacency matrix is necessary to achieve the diversity of signals necessary for low training error. We test this assumption by varying both sparsity, defined as the fraction of network edges that are not zero, and the fraction of edges flipped $\varepsilon_f$. Sparsity will be denoted as $\phi$. 
 
\begin{figure}[ht]
\centering
\includegraphics[scale=0.6]{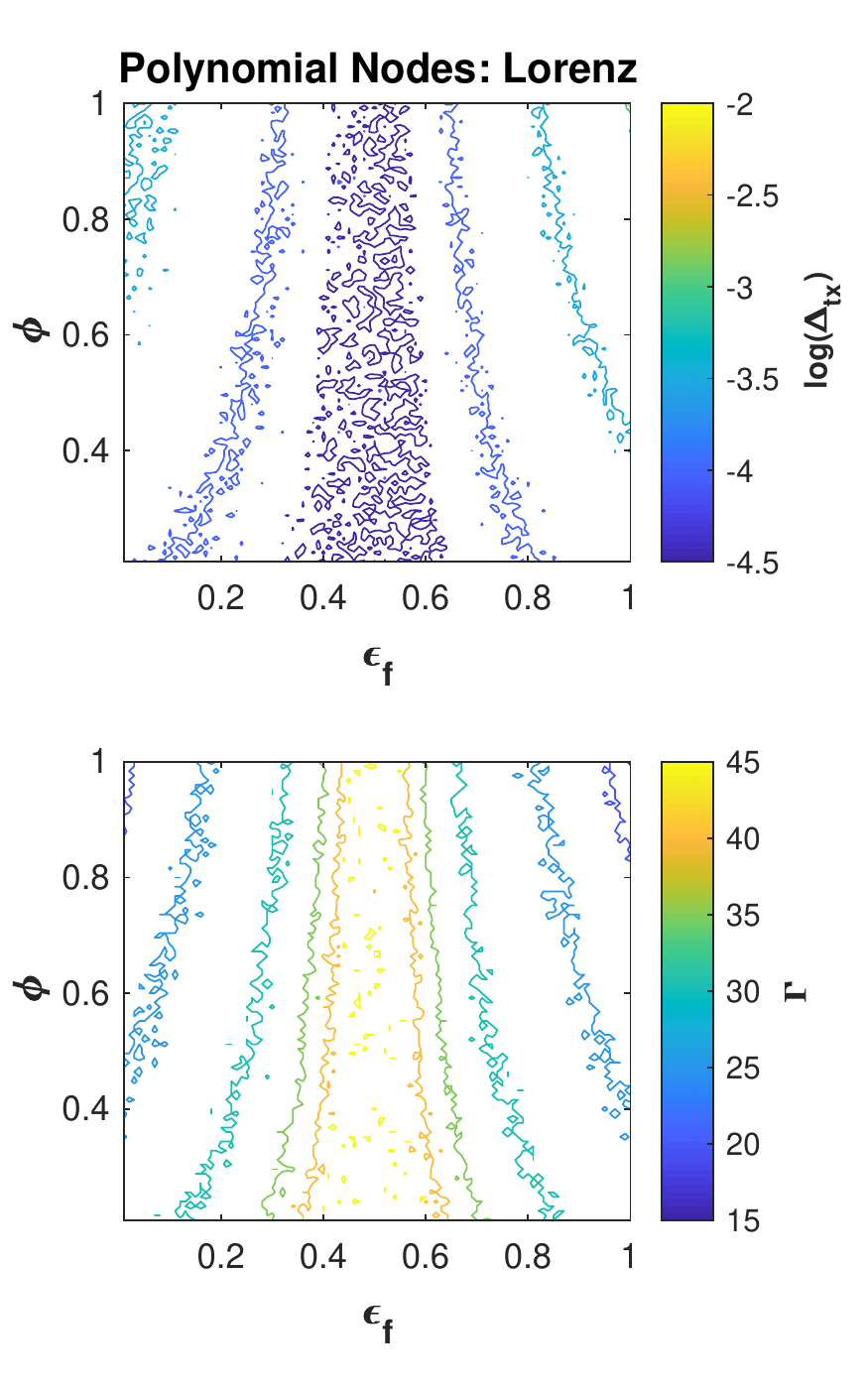} 
  \caption{ \label{nleq_lor_flipsparse} Top contour plot is the log base 10 of the training error $\Delta_{tx}$ for the polynomial nodes driven by the Lorenz $x$ signal. The horizontal axis is the fraction of edges flipped, $\varepsilon_f$, while the vertical axis is the sparsity $\phi$, which is equal to the fraction of nonzero edges. The lower plot is the covariance rank $\Gamma$.
}
  \end{figure}

\begin{figure}[ht]
\centering
\includegraphics[scale=0.6]{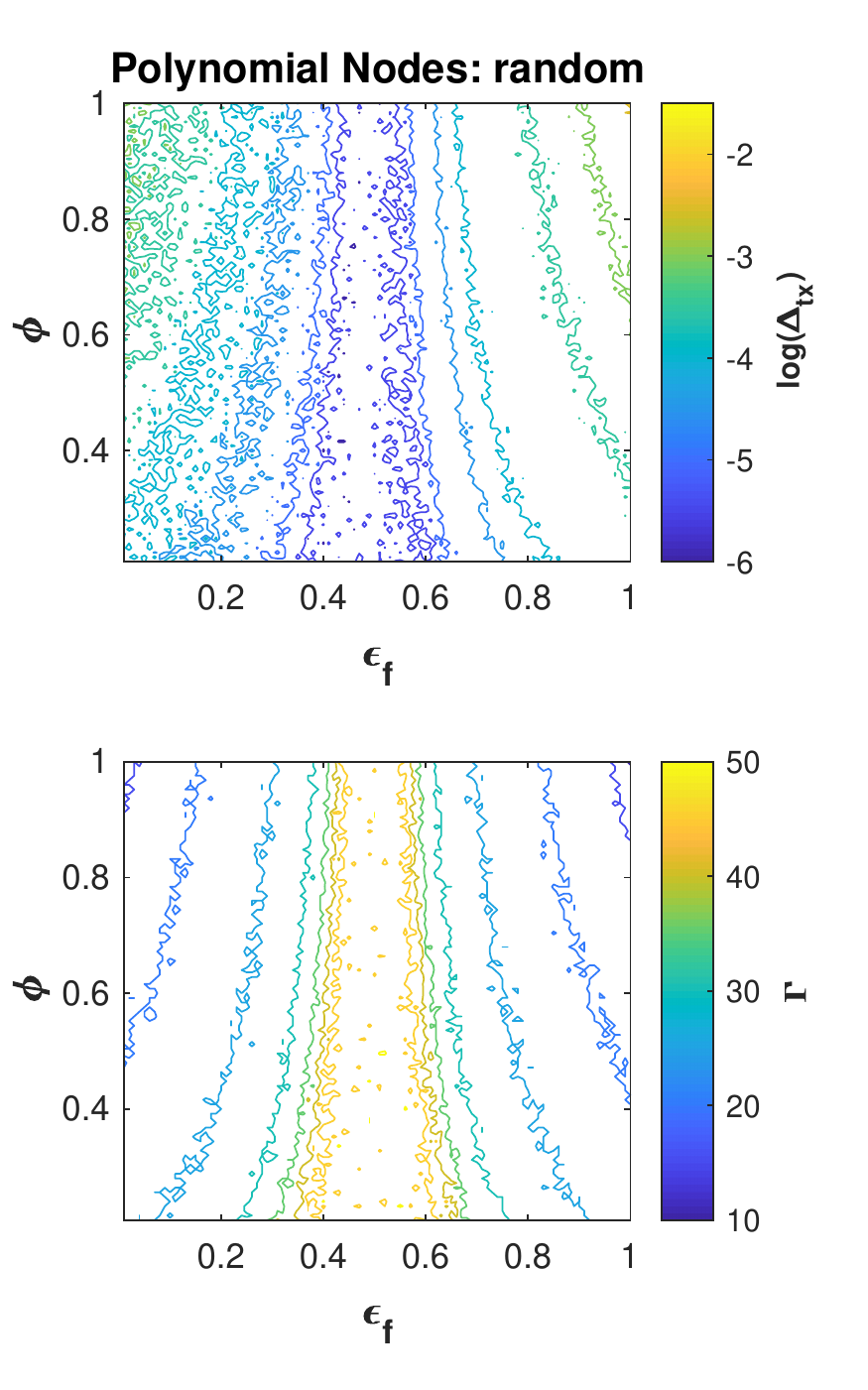} 
  \caption{ \label{nleq_narma_flipsparse} Top contour plot is the log base 10 of the training error $\Delta_{tx}$ for the polynomial nodes driven by the random signal $x(k)$ from eq. (\ref{narma}). The horizontal axis is the fraction of edges flipped, $\varepsilon_f$, while the vertical axis is the sparsity $\phi$, which is equal to the fraction of nonzero edges. The lower plot is the covariance rank $\Gamma$.
}
  \end{figure}

 The top plot in figure \ref{nleq_lor_flipsparse} is a contour plot of the log base 10 testing error $\Delta_{tx}$ as the fraction of edges flipped and the sparsity are varied. The lower plot is the covariance rank $\Gamma$. The plots in figure \ref{nleq_lor_flipsparse} show that the testing error gets smaller and covariance rank gets larger as the sparsity decreases, so choosing sparse networks can be a useful goal. These plots do show, however, that the influence of the number of edges flipped is much stronger. Figure \ref{nleq_narma_flipsparse} shows the same information for the polynomial nodes driven by the random signal $x(k)$ from eq. (\ref{narma}).
  
     \begin{figure}[ht]
\centering
\includegraphics[scale=0.6]{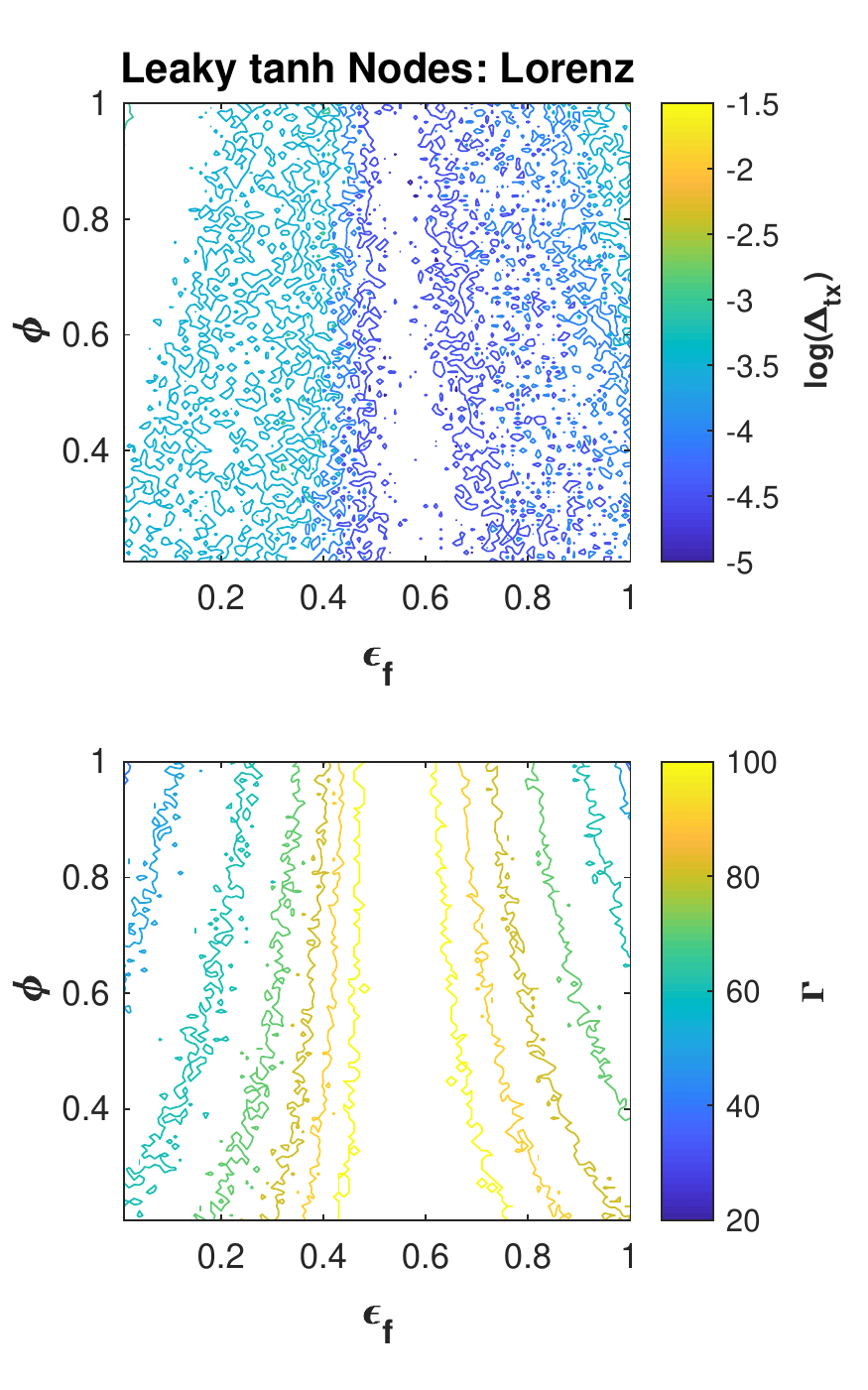} 
  \caption{ \label{umd_lorenz_flipsparse} Top contour plot is the log base 10 of the training error $\Delta_{tx}$ for the leaky tanh nodes driven by the Lorenz $x$ signal. The horizontal axis is the fraction of edges flipped, $\varepsilon_f$, while the vertical axis is the sparsity $\phi$, which is equal to the fraction of nonzero edges. The lower plot is the covariance rank $\Gamma$.
}
  \end{figure} 
 
     \begin{figure}[ht]
\centering
\includegraphics[scale=0.6]{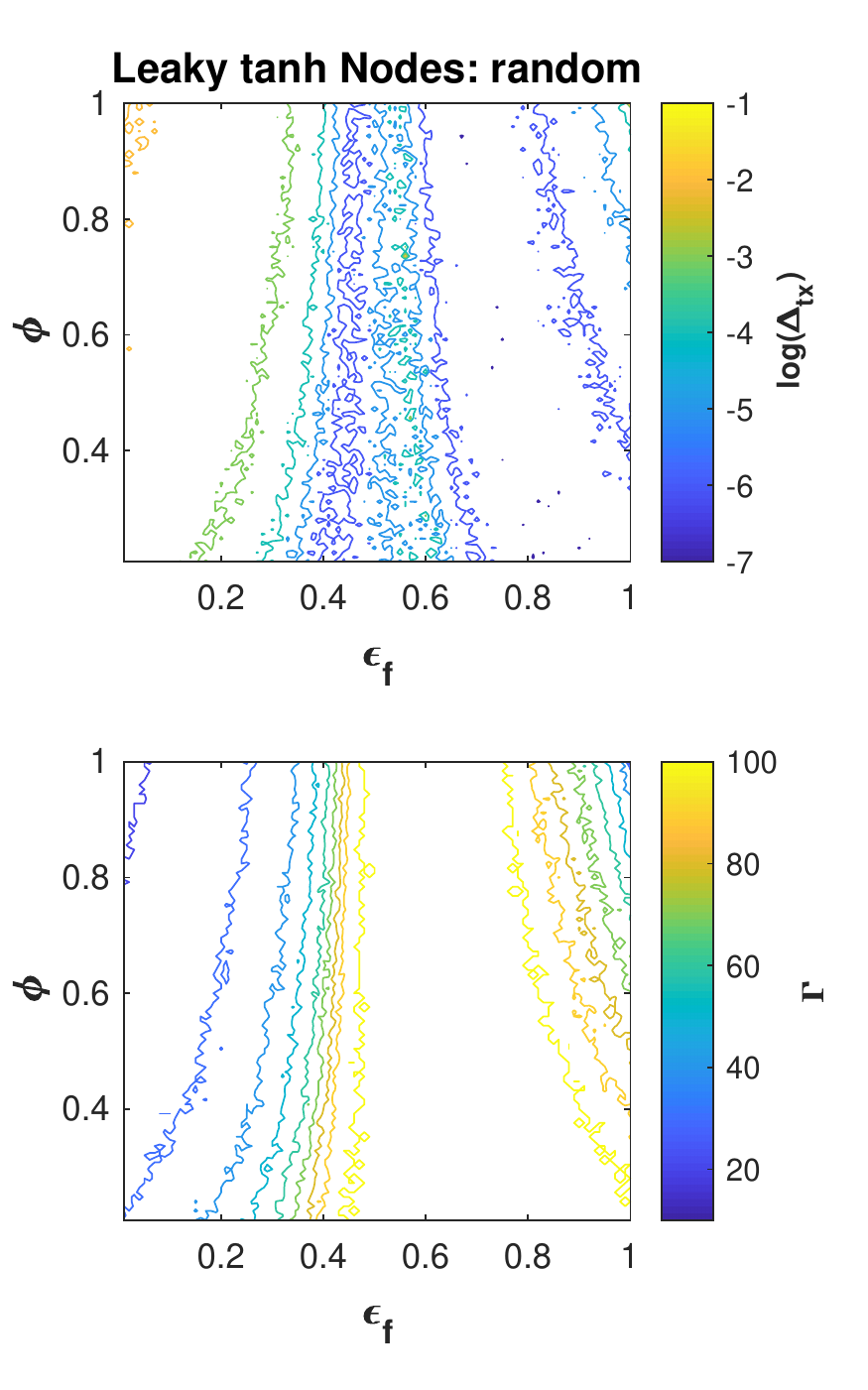} 
  \caption{ \label{umd_narma_flipsparse} Top contour plot is the log base 10 of the training error $\Delta_{tx}$ for the leaky tanh nodes driven by the random signal $x(k)$ from eq. (\ref{narma}). The horizontal axis is the fraction of edges flipped, $\varepsilon_f$, while the vertical axis is the sparsity $\phi$, which is equal to the fraction of nonzero edges. The lower plot is the covariance rank $\Gamma$.
}
  \end{figure}
  
    Figures \ref{umd_lorenz_flipsparse} and \ref{umd_narma_flipsparse} are contour plots for a network of leaky tanh nodes being driven by the Lorenz $x$ signal or the random signal $x(k)$ from eq. (\ref{narma}). These two figures show an asymmetry along the $\varepsilon_f$ axis, especially for the random driving signal. The equation for the leaky tanh nodes contains an offset of 1.0, so the equations are not symmetric about zero. Changing most of the network edges from +1 to -1 can produce a bias in the network signals, which will affect the dynamics because of the asymmetry in the leaky tanh node equation. 
    
\section{Conclusions}
We have simulated reservoir computers where the edges between the reservoir computer nodes were all +1 or 0, and then changed the reservoir computer network by flipping some of these edges from +1 to -1. We have done this for different combinations of node type and input signals.

If a small fraction of the edges of the adjacency matrix were flipped from +1 to -1, the number of symmetries in the adjacency matrix could function as a similarity measure for the adjacency matrix. Adjacency matrices that had more symmetries led to reservoirs that had a lower covariance rank and larger errors in fitting a training signal. 

When the only symmetry in the adjacency matrix was the identity, a different measure of the variation in the adjacency matrix was necessary. Because the adjacency matrices used in this work were simple, we could use the fraction $\varepsilon_f$ of elements flipped from +1 to -1 as a measure of this variation. Increasing $\varepsilon_f$ increased the rank of the covariance of the adjacency matrix, which in most cases led to a smaller error in fitting a training signal. Comparing to completely random networks, we found that our networks, with all edges $\pm 1$, produced as small of a testing error as the completely random network,

We also investigated the relation between the fraction of edges flipped and memory capacity. We quantified memory capacity using the method of \cite{jaeger2002}, where memory is measured by finding how well the reservoir can fit previous values of a random input signal. We found that within a particular node type, memory capacity, testing error and covariance rank were all correlated.

Studying testing error and covariance rank showed that having fewer nonzero edges in the network (lower sparsity) did produce a smaller testing error, but the effect of sparsity was not as strong as the effect of flipping more network edges.

We did not investigate the effect of different nonrandom network statistics on the behavior of the reservoir computer. Networks whose connections are not all $\pm 1$ will require different measures of diversity. There are other types of network statistics such as nearest neighbor networks, star, or ring networks, or networks with different weights for different connections.  All of these types of networks may perform differently as their parameters are changed.

\pagebreak

\section{References}
\bibliography{network_structure}{}

\begin{thebibliography}{32}
\expandafter\ifx\csname natexlab\endcsname\relax\def\natexlab#1{#1}\fi
\expandafter\ifx\csname bibnamefont\endcsname\relax
  \def\bibnamefont#1{#1}\fi
\expandafter\ifx\csname bibfnamefont\endcsname\relax
  \def\bibfnamefont#1{#1}\fi
\expandafter\ifx\csname citenamefont\endcsname\relax
  \def\citenamefont#1{#1}\fi
\expandafter\ifx\csname url\endcsname\relax
  \def\url#1{\texttt{#1}}\fi
\expandafter\ifx\csname urlprefix\endcsname\relax\def\urlprefix{URL }\fi
\providecommand{\bibinfo}[2]{#2}
\providecommand{\eprint}[2][]{\url{#2}}

\bibitem[{\citenamefont{Jaeger}(2001)}]{jaeger2001}
\bibinfo{author}{\bibfnamefont{H.}~\bibnamefont{Jaeger}},
  \bibinfo{journal}{German National Research Center for Information Technology
  GMD Technical Report} \textbf{\bibinfo{volume}{148}}, \bibinfo{pages}{34}
  (\bibinfo{year}{2001}).

\bibitem[{\citenamefont{Natschlaeger et~al.}(2002)\citenamefont{Natschlaeger,
  Maass, and Markram}}]{natschlaeger2002}
\bibinfo{author}{\bibfnamefont{T.}~\bibnamefont{Natschlaeger}},
  \bibinfo{author}{\bibfnamefont{W.}~\bibnamefont{Maass}}, \bibnamefont{and}
  \bibinfo{author}{\bibfnamefont{H.}~\bibnamefont{Markram}},
  \bibinfo{journal}{Special Issue on Foundations of Information Processing of
  TELEMATIK} \textbf{\bibinfo{volume}{8}}, \bibinfo{pages}{39}
  (\bibinfo{year}{2002}).

\bibitem[{\citenamefont{Lu et~al.}(2018)\citenamefont{Lu, Hunt, and
  Ott}}]{lu2018}
\bibinfo{author}{\bibfnamefont{Z.}~\bibnamefont{Lu}},
  \bibinfo{author}{\bibfnamefont{B.~R.} \bibnamefont{Hunt}}, \bibnamefont{and}
  \bibinfo{author}{\bibfnamefont{E.}~\bibnamefont{Ott}},
  \bibinfo{journal}{Chaos: An Interdisciplinary Journal of Nonlinear Science}
  \textbf{\bibinfo{volume}{28}}, \bibinfo{pages}{061104}
  (\bibinfo{year}{2018}).

\bibitem[{\citenamefont{Zimmermann and Parlitz}(2018)}]{zimmerman2018}
\bibinfo{author}{\bibfnamefont{R.~S.} \bibnamefont{Zimmermann}}
  \bibnamefont{and} \bibinfo{author}{\bibfnamefont{U.}~\bibnamefont{Parlitz}},
  \bibinfo{journal}{Chaos: An Interdisciplinary Journal of Nonlinear Science}
  \textbf{\bibinfo{volume}{28}}, \bibinfo{pages}{043118}
  (\bibinfo{year}{2018}).

\bibitem[{\citenamefont{Antonik et~al.}(2018)\citenamefont{Antonik, Gulina,
  Pauwels, and Massar}}]{antonik2018}
\bibinfo{author}{\bibfnamefont{P.}~\bibnamefont{Antonik}},
  \bibinfo{author}{\bibfnamefont{M.}~\bibnamefont{Gulina}},
  \bibinfo{author}{\bibfnamefont{J.}~\bibnamefont{Pauwels}}, \bibnamefont{and}
  \bibinfo{author}{\bibfnamefont{S.}~\bibnamefont{Massar}},
  \bibinfo{journal}{Physical Review E} \textbf{\bibinfo{volume}{98}},
  \bibinfo{pages}{012215} (\bibinfo{year}{2018}).

\bibitem[{\citenamefont{Lu et~al.}(2017)\citenamefont{Lu, Pathak, Hunt, Girvan,
  Brockett, and Ott}}]{lu2017}
\bibinfo{author}{\bibfnamefont{Z.}~\bibnamefont{Lu}},
  \bibinfo{author}{\bibfnamefont{J.}~\bibnamefont{Pathak}},
  \bibinfo{author}{\bibfnamefont{B.}~\bibnamefont{Hunt}},
  \bibinfo{author}{\bibfnamefont{M.}~\bibnamefont{Girvan}},
  \bibinfo{author}{\bibfnamefont{R.}~\bibnamefont{Brockett}}, \bibnamefont{and}
  \bibinfo{author}{\bibfnamefont{E.}~\bibnamefont{Ott}},
  \bibinfo{journal}{Chaos: An Interdisciplinary Journal of Nonlinear Science}
  \textbf{\bibinfo{volume}{27}}, \bibinfo{pages}{041102}
  (\bibinfo{year}{2017}).

\bibitem[{\citenamefont{Jaeger and Haas}(2004)}]{jaeger2004}
\bibinfo{author}{\bibfnamefont{H.}~\bibnamefont{Jaeger}} \bibnamefont{and}
  \bibinfo{author}{\bibfnamefont{H.}~\bibnamefont{Haas}},
  \bibinfo{journal}{Science} \textbf{\bibinfo{volume}{304}},
  \bibinfo{pages}{78} (\bibinfo{year}{2004}).

\bibitem[{\citenamefont{Jalalvand et~al.}(2018)\citenamefont{Jalalvand,
  Demuynck, Neve, and Martens}}]{jalavand2018}
\bibinfo{author}{\bibfnamefont{A.}~\bibnamefont{Jalalvand}},
  \bibinfo{author}{\bibfnamefont{K.}~\bibnamefont{Demuynck}},
  \bibinfo{author}{\bibfnamefont{W.~D.} \bibnamefont{Neve}}, \bibnamefont{and}
  \bibinfo{author}{\bibfnamefont{J.-P.} \bibnamefont{Martens}},
  \bibinfo{journal}{Neurocomputing} \textbf{\bibinfo{volume}{277}},
  \bibinfo{pages}{237} (\bibinfo{year}{2018}).

\bibitem[{\citenamefont{Luko{\v s}evi{\v c}ius
  et~al.}(2012)\citenamefont{Luko{\v s}evi{\v c}ius, Jaeger, and
  Schrauwen}}]{lukosevicius2012}
\bibinfo{author}{\bibfnamefont{M.}~\bibnamefont{Luko{\v s}evi{\v c}ius}},
  \bibinfo{author}{\bibfnamefont{H.}~\bibnamefont{Jaeger}}, \bibnamefont{and}
  \bibinfo{author}{\bibfnamefont{B.}~\bibnamefont{Schrauwen}},
  \bibinfo{journal}{KI - Künstliche Intelligenz} \textbf{\bibinfo{volume}{26}},
  \bibinfo{pages}{365} (\bibinfo{year}{2012}).

\bibitem[{\citenamefont{Larger et~al.}(2012)\citenamefont{Larger, Soriano,
  Brunner, Appeltant, Gutierrez, Pesquera, Mirasso, and Fischer}}]{larger2012}
\bibinfo{author}{\bibfnamefont{L.}~\bibnamefont{Larger}},
  \bibinfo{author}{\bibfnamefont{M.~C.} \bibnamefont{Soriano}},
  \bibinfo{author}{\bibfnamefont{D.}~\bibnamefont{Brunner}},
  \bibinfo{author}{\bibfnamefont{L.}~\bibnamefont{Appeltant}},
  \bibinfo{author}{\bibfnamefont{J.~M.} \bibnamefont{Gutierrez}},
  \bibinfo{author}{\bibfnamefont{L.}~\bibnamefont{Pesquera}},
  \bibinfo{author}{\bibfnamefont{C.~R.} \bibnamefont{Mirasso}},
  \bibnamefont{and} \bibinfo{author}{\bibfnamefont{I.}~\bibnamefont{Fischer}},
  \bibinfo{journal}{Optics Express} \textbf{\bibinfo{volume}{20}},
  \bibinfo{pages}{3241} (\bibinfo{year}{2012}).

\bibitem[{\citenamefont{der Sande et~al.}(2017)\citenamefont{der Sande,
  Brunner, and Soriano}}]{van_der_sande2017}
\bibinfo{author}{\bibfnamefont{G.~V.} \bibnamefont{der Sande}},
  \bibinfo{author}{\bibfnamefont{D.}~\bibnamefont{Brunner}}, \bibnamefont{and}
  \bibinfo{author}{\bibfnamefont{M.~C.} \bibnamefont{Soriano}},
  \bibinfo{journal}{Nanophotonics} \textbf{\bibinfo{volume}{6}},
  \bibinfo{pages}{561} (\bibinfo{year}{2017}).

\bibitem[{\citenamefont{Schurmann et~al.}(2004)\citenamefont{Schurmann, Meier,
  and Schemmel}}]{schurmann2004}
\bibinfo{author}{\bibfnamefont{F.}~\bibnamefont{Schurmann}},
  \bibinfo{author}{\bibfnamefont{K.}~\bibnamefont{Meier}}, \bibnamefont{and}
  \bibinfo{author}{\bibfnamefont{J.}~\bibnamefont{Schemmel}}, in
  \emph{\bibinfo{booktitle}{Advances in Neural Information Processing Systems
  17}} (\bibinfo{publisher}{MIT Press}, \bibinfo{year}{2004}), pp.
  \bibinfo{pages}{1201--1208}.

\bibitem[{\citenamefont{Dion et~al.}(2018)\citenamefont{Dion, Mejaouri, and
  Sylvestre}}]{dion2018}
\bibinfo{author}{\bibfnamefont{G.}~\bibnamefont{Dion}},
  \bibinfo{author}{\bibfnamefont{S.}~\bibnamefont{Mejaouri}}, \bibnamefont{and}
  \bibinfo{author}{\bibfnamefont{J.}~\bibnamefont{Sylvestre}},
  \bibinfo{journal}{Journal of Applied Physics} \textbf{\bibinfo{volume}{124}},
  \bibinfo{pages}{152132} (\bibinfo{year}{2018}).

\bibitem[{\citenamefont{Canaday et~al.}(2018)\citenamefont{Canaday, Griffith,
  and Gauthier}}]{canaday2018}
\bibinfo{author}{\bibfnamefont{D.}~\bibnamefont{Canaday}},
  \bibinfo{author}{\bibfnamefont{A.}~\bibnamefont{Griffith}}, \bibnamefont{and}
  \bibinfo{author}{\bibfnamefont{D.~J.} \bibnamefont{Gauthier}},
  \bibinfo{journal}{Chaos: An Interdisciplinary Journal of Nonlinear Science}
  \textbf{\bibinfo{volume}{28}}, \bibinfo{pages}{123119}
  (\bibinfo{year}{2018}).

\bibitem[{\citenamefont{Jaeger}(2002)}]{jaeger2002}
\bibinfo{author}{\bibfnamefont{H.}~\bibnamefont{Jaeger}},
  \bibinfo{journal}{Technical report GMD-Forschungszentrum Informationstechnik}
   (\bibinfo{year}{2002}).

\bibitem[{\citenamefont{Inubushi and Yoshimura}(2017)}]{inubushi2017}
\bibinfo{author}{\bibfnamefont{M.}~\bibnamefont{Inubushi}} \bibnamefont{and}
  \bibinfo{author}{\bibfnamefont{K.}~\bibnamefont{Yoshimura}},
  \bibinfo{journal}{Scientific Reports} \textbf{\bibinfo{volume}{7}},
  \bibinfo{pages}{10199} (\bibinfo{year}{2017}).

\bibitem[{\citenamefont{Marzen}(2017)}]{marzen2017}
\bibinfo{author}{\bibfnamefont{S.}~\bibnamefont{Marzen}},
  \bibinfo{journal}{Physical Review E} \textbf{\bibinfo{volume}{96}},
  \bibinfo{pages}{032308} (\bibinfo{year}{2017}).

\bibitem[{\citenamefont{Lymburn et~al.}(2019)\citenamefont{Lymburn, Khor,
  Stemler, Corrêa, Small, and Jüngling}}]{lymburn2019}
\bibinfo{author}{\bibfnamefont{T.}~\bibnamefont{Lymburn}},
  \bibinfo{author}{\bibfnamefont{A.}~\bibnamefont{Khor}},
  \bibinfo{author}{\bibfnamefont{T.}~\bibnamefont{Stemler}},
  \bibinfo{author}{\bibfnamefont{D.~C.} \bibnamefont{Corrêa}},
  \bibinfo{author}{\bibfnamefont{M.}~\bibnamefont{Small}}, \bibnamefont{and}
  \bibinfo{author}{\bibfnamefont{T.}~\bibnamefont{Jüngling}},
  \bibinfo{journal}{Chaos: An Interdisciplinary Journal of Nonlinear Science}
  \textbf{\bibinfo{volume}{29}}, \bibinfo{pages}{023118}
  (\bibinfo{year}{2019}).

\bibitem[{\citenamefont{Rodan and Tino}(2011)}]{rodan2011}
\bibinfo{author}{\bibfnamefont{A.}~\bibnamefont{Rodan}} \bibnamefont{and}
  \bibinfo{author}{\bibfnamefont{P.}~\bibnamefont{Tino}},
  \bibinfo{journal}{IEEE Transactions on Neural Networks}
  \textbf{\bibinfo{volume}{22}}, \bibinfo{pages}{131} (\bibinfo{year}{2011}).

\bibitem[{\citenamefont{Manjunath and Jaeger}(2013)}]{manjunath2013}
\bibinfo{author}{\bibfnamefont{G.}~\bibnamefont{Manjunath}} \bibnamefont{and}
  \bibinfo{author}{\bibfnamefont{H.}~\bibnamefont{Jaeger}},
  \bibinfo{journal}{Neural Computation} \textbf{\bibinfo{volume}{25}},
  \bibinfo{pages}{671} (\bibinfo{year}{2013}).

\bibitem[{\citenamefont{Verstraeten et~al.}(2010)\citenamefont{Verstraeten,
  Dambre, Dutoit, and Schrauwen}}]{verstraeten2010}
\bibinfo{author}{\bibfnamefont{D.}~\bibnamefont{Verstraeten}},
  \bibinfo{author}{\bibfnamefont{J.}~\bibnamefont{Dambre}},
  \bibinfo{author}{\bibfnamefont{X.}~\bibnamefont{Dutoit}}, \bibnamefont{and}
  \bibinfo{author}{\bibfnamefont{B.}~\bibnamefont{Schrauwen}}, in
  \emph{\bibinfo{booktitle}{The 2010 International Joint Conference on Neural
  Networks (IJCNN)}} (\bibinfo{publisher}{IEEE}, \bibinfo{year}{2010}), pp.
  \bibinfo{pages}{1--8}.

\bibitem[{\citenamefont{Hadaeghi and Jaeger}(2019)}]{hadaeghi2019}
\bibinfo{author}{\bibfnamefont{F.}~\bibnamefont{Hadaeghi}} \bibnamefont{and}
  \bibinfo{author}{\bibfnamefont{H.}~\bibnamefont{Jaeger}},
  \bibinfo{journal}{Neurocomputing}  (\bibinfo{year}{2019}).

\bibitem[{\citenamefont{Penrose}(1955)}]{penrose1955}
\bibinfo{author}{\bibfnamefont{R.}~\bibnamefont{Penrose}},
  \bibinfo{journal}{Mathematical Proceedings of the Cambridge Philosophical
  Society} \textbf{\bibinfo{volume}{51}}, \bibinfo{pages}{406}
  (\bibinfo{year}{1955}).

\bibitem[{\citenamefont{Tikhonov}(1943)}]{tikhonov1943}
\bibinfo{author}{\bibfnamefont{A.~N.} \bibnamefont{Tikhonov}},
  \bibinfo{journal}{Comptes Rendus De L Academie Des Sciences De L Urss}
  \textbf{\bibinfo{volume}{39}}, \bibinfo{pages}{176} (\bibinfo{year}{1943}).

\bibitem[{\citenamefont{Golub et~al.}(1979)\citenamefont{Golub, Heath, and
  Wahba}}]{golub1979}
\bibinfo{author}{\bibfnamefont{G.~H.} \bibnamefont{Golub}},
  \bibinfo{author}{\bibfnamefont{M.}~\bibnamefont{Heath}}, \bibnamefont{and}
  \bibinfo{author}{\bibfnamefont{G.}~\bibnamefont{Wahba}},
  \bibinfo{journal}{Technometrics} \textbf{\bibinfo{volume}{21}},
  \bibinfo{pages}{215} (\bibinfo{year}{1979}).

\bibitem[{\citenamefont{Golubitsky et~al.}(1985)\citenamefont{Golubitsky,
  Stewart, and Schaeffer}}]{GolubitskyBOOKII}
\bibinfo{author}{\bibfnamefont{M.}~\bibnamefont{Golubitsky}},
  \bibinfo{author}{\bibfnamefont{I.}~\bibnamefont{Stewart}}, \bibnamefont{and}
  \bibinfo{author}{\bibfnamefont{D.}~\bibnamefont{Schaeffer}},
  \emph{\bibinfo{title}{Singularities and groups in bifurcation theory}},
  vol.~\bibinfo{volume}{II} (\bibinfo{publisher}{Springer-Verlag},
  \bibinfo{address}{New York, NY}, \bibinfo{year}{1985}).

\bibitem[{\citenamefont{Stein}(2013)}]{Stein}
\bibinfo{author}{\bibfnamefont{W.}~\bibnamefont{Stein}},
  \emph{\bibinfo{title}{SAGE: Software for Algebra and Geometry
  Experimentation}} (\bibinfo{publisher}{http://www.sagemath.org/sage/,
  http://sage.scipy.org/}, \bibinfo{year}{2013}).

\bibitem[{\citenamefont{Pecora et~al.}(2014)\citenamefont{Pecora, Sorrentino,
  Hagerstrom, Murphy, and Roy}}]{PecoraClusterSyncNat2014}
\bibinfo{author}{\bibfnamefont{L.~M.} \bibnamefont{Pecora}},
  \bibinfo{author}{\bibfnamefont{F.}~\bibnamefont{Sorrentino}},
  \bibinfo{author}{\bibfnamefont{A.~M.} \bibnamefont{Hagerstrom}},
  \bibinfo{author}{\bibfnamefont{T.~E.} \bibnamefont{Murphy}},
  \bibnamefont{and} \bibinfo{author}{\bibfnamefont{R.}~\bibnamefont{Roy}},
  \bibinfo{journal}{Nature communications} \textbf{\bibinfo{volume}{5}}
  (\bibinfo{year}{2014}).

\bibitem[{\citenamefont{Cho et~al.}(2017)\citenamefont{Cho, Nishikawa, and
  Motter}}]{ChoNishikMotterPRL2017}
\bibinfo{author}{\bibfnamefont{Y.}~\bibnamefont{Cho}},
  \bibinfo{author}{\bibfnamefont{T.}~\bibnamefont{Nishikawa}},
  \bibnamefont{and} \bibinfo{author}{\bibfnamefont{A.}~\bibnamefont{Motter}},
  \bibinfo{journal}{Phys. Rev. Lett.} \textbf{\bibinfo{volume}{119}},
  \bibinfo{pages}{084101} (\bibinfo{year}{2017}).

\bibitem[{\citenamefont{Sorrentino and
  Pecora}(2016)}]{SorrentinoApproxSymm2016}
\bibinfo{author}{\bibfnamefont{F.}~\bibnamefont{Sorrentino}} \bibnamefont{and}
  \bibinfo{author}{\bibfnamefont{L.}~\bibnamefont{Pecora}},
  \bibinfo{journal}{CHAOS} \textbf{\bibinfo{volume}{26}},
  \bibinfo{pages}{094823} (\bibinfo{year}{2016}).

\bibitem[{\citenamefont{Jolliffe}(2011)}]{joliffe2011}
\bibinfo{author}{\bibfnamefont{I.~T.} \bibnamefont{Jolliffe}},
  \emph{\bibinfo{title}{Principal component analysis}}
  (\bibinfo{publisher}{Springer}, \bibinfo{year}{2011}).

\bibitem[{\citenamefont{Lorenz}(1963)}]{lorenz1963}
\bibinfo{author}{\bibfnamefont{E.~N.} \bibnamefont{Lorenz}},
  \bibinfo{journal}{Journal of Atmospheric Science}
  \textbf{\bibinfo{volume}{20}}, \bibinfo{pages}{130} (\bibinfo{year}{1963}).

\end{thebibliography}

\end{document}